\documentclass[a4paper,11pt]{article}
\pdfoutput=1 % if your are submitting a pdflatex (i.e. if you have
\usepackage{jheppub} % for details on the use of the package, please
\usepackage[T1]{fontenc} % if needed
\usepackage[utf8]{inputenc}                 % tastiera italiana \, per spazio piccolo, \! spazio negativo, ~ per spazio fisso e \LaTeX\ per chiudere
\usepackage{microtype}                      % migliora riempimento righe
\usepackage{lmodern}

\usepackage{amssymb}                % simboli matematica
\usepackage{amsmath}                % formule e simboli matematica, PACCHETTO PRINCIPALE
\usepackage{amsthm}                 % teoremi, dimostrazioni, etc... comando \newtheorem
\usepackage{mathrsfs}               % stili di font speciali in ambiente matematico (fraktur: mathfrak, blackboard: mathbb)
\usepackage{mathtools}              % per definizione di nuovi strumenti matematici, come DeclarePairedDelimiter
\usepackage{comment}
\usepackage{appendix}
\usepackage{slashed}
\usepackage{dsfont}
\usepackage{xcolor}

\newcommand*\dif{\mathop{}\!\mathrm{d}}             % la d per indicare il differenziale negli integrali
\newcommand{\vect}[1]{\boldsymbol{\mathbf{#1}}}     % vettori con lettere in grassetto

\let\temp\varepsilon                                % inverto definizione di \epsilon e \varepsilon
\let\varepsilon\epsilon
\let\epsilon\temp

\makeatletter
\def\@fpheader{\relax}
\makeatother

\title{Relaxation terms for anomalous hydrodynamic transport in Weyl semimetals from kinetic theory}

\author[1,2]{Andrea Amoretti,}
\author[3]{Daniel K. Brattan,}
\author[1,2]{Luca Martinoia,}
\author[1,2]{Ioannis Matthaiakakis}
\author[1,2]{and Jonas Rongen}

\affiliation[1]{Dipartimento di Fisica, Universit\`a di Genova, via Dodecaneso 33, I-16146, Genova, Italy}
\affiliation[2]{I.N.F.N. - Sezione di Genova, via Dodecaneso 33, I-16146, Genova, Italy}
\affiliation[3]{CPHT, \'{E}cole Polytechnique, 91128 Palaiseau cedex, France}

\emailAdd{andrea.amoretti@ge.infn.it}
\emailAdd{danny.brattan@gmail.com}
\emailAdd{luca.martinoia@ge.infn.it}
\emailAdd{giannismattheakakis@gmail.com}
\emailAdd{jonas.ludovico.rongen@edu.unige.it}

\preprint{CPHT-RR060.092023}

\abstract{We consider as a model of Weyl semimetal thermoelectric transport a $(3+1)$-dimensional charged, relativistic and relaxed fluid with a $U(1)_{V} \times U(1)_{A}$ chiral anomaly. We take into account all possible mixed energy, momentum, electric and chiral charge relaxations, and discover which are compatible with electric charge conservation, Onsager reciprocity and a finite DC conductivity. We find that all relaxations respecting these constraints necessarily render the system open and violate the second law of thermodynamics. We then demonstrate how the relaxations we have found arise from kinetic theory and a modified relaxation time approximation. Our results lead to DC conductivities that differ from those found in the literature opening the path to experimental verification.}

\notoc

\begin{document} 
\maketitle
\flushbottom
\newpage

\section{Introduction}

In recent years there has been an increased interest in the physics of semimetals and particularly semimetals where the valence and conduction bands have a linear crossing point. We focus our attention on Weyl semimetals (WSM) which are a class of quantum materials hosting Weyl fermions \cite{armitage:weylanddirac,burkov:weylmetals,Hosur_2013}. These materials enjoy an accidental chiral symmetry because the band-touching points (the Weyl nodes) must always come in pairs of opposite chirality \cite{Nielsen:1980rz,NIELSEN1981219}. The nodes will generically lie on top of each other in momentum space, in which case we speak of a Dirac semimetal, unless the underlying electronic system breaks either time-reversal or parity.  We are interested in these kinds of materials, because their low-energy excitations can be described by relativistic quantum field theories typically used in high energy physics. As such they allow for tabletop experimental probes of high-energy phenomena. 

The high-energy phenomenon we are interested in is the $U(1)_A\times U(1)_V$ anomaly of massless relativistic fermions \cite{PhysRev.177.2426,Bell1969,nielsen:adler-bell-jackiw}. The anomaly is generated by co-linear electromagnetic fields and, hence, strongly affects thermoelectric transport in WSMs \cite{nielsen:adler-bell-jackiw, landsteiner:notesanomaly,landsteiner:negativemagnetoresistivity, landsteiner:anomaloustransport, chernodub:thermaltransport, doi:10.1080/23746149.2017.1327329, sukhachov:anomalousgurzhi, lucas:hydrodynamictheory, abbasi:magnetotransportanomalous, abbasi:magnetotransportchiral, gorbar:nonlocaltransport, son:chiralanomaly,Nag_2021}. More precisely, we are interested in the anomaly's effect on the negative magneto-resistance (NMR) of WSMs in the hydrodynamic regime \cite{landsteiner:negativemagnetoresistivity,lucas:hydrodynamictheory,abbasi:magnetotransportanomalous, chernodub:thermaltransport}. The signature effect of the anomaly on the NMR is the $B^2$ growth of thermoelectric conductivities, e.g. for the electric conductivity $\sigma\simeq \sigma_\text{Drude}+\alpha B^2$ at small $B$. Hydrodynamic behaviour in WSMs has been observed in \cite{Gooth_2018}, while many experiments found NMR effects in electric and thermal transport compatible with the axial anomaly \cite{Vu:thermal_chiral_anomaly,Xiong:evidence_chiral_anomaly,PhysRevX.5.031023,Gooth_2017,Jia2016ThermoelectricSO}.\footnote{We use the term ``chiral'' interchangeably with ``axial'' throughout the paper.}

Although we have an excellent description of the NMR at small magnetic fields, there are some open issues we need to address. Firstly, it was recently shown that changing the hydrodynamic frame does change the explicit formulas for the NMR contribution to the conductivities often found in the literature \cite{amoretti:framedependence}. In addition, some of the models used have conductivities that are not Onsager reciprocal and thus break microscopic time-reversal invariance---even if we use the prescription advocated in \cite{amoretti:framedependence}. While there are models where the conductivities are Onsager reciprocal, they require that the WSM fluid loses energy, electric and chiral charge at the same rate. This contradicts experimental expectations in condensed matter systems, where energy and electric charge are expected to be almost identically conserved, while chiral charge is expected to relax quickly to its equilibrium value.

In this work we show that both of these issues can be circumvented. In particular, we avoid the hydrodynamic frame issue altogether by studying the NMR using the zeroth order hydrodynamics of an anomalous fluid \cite{amoretti:framedependence}. Then, we show that it is possible to decouple the relaxation rates for energy, electric and chiral charge, while still satisfying Onsager reciprocity. We achieve this by introducing generalized relaxations that couple these charges as in \cite{amoretti:restoringtime}. Not all of the generalized relaxations are independent, instead they fall into families of Onsager-reciprocal solutions that depend on at most three free parameters. Importantly, the momentum relaxation rate can remain independent of all other relaxation times---as expected---only if we relax the normal, non-anomalous charges of the fluid. This can be justified by observing that the anomalous component of the fluid resembles a superfluid, i.e. it does not produce any drag or entropy, and thus should not relax. This is what is usually done in the context of kinetic theory \cite{gorbar:consistenthydrodynamics,dantas:magnetotransportmultiweyl} and also in \cite{lucas:hydrodynamictheory}, albeit the momentum relaxation is not implemented in terms of a relaxation rate, but via a disordered chemical potential. 

Moreover, we supplement our phenomenological hydrodynamic approach via a kinetic theory derivation of generalized relaxation times. Namely, we develop a novel relaxation time approximation (RTA) for the electron-electron, electron-phonon and electron-impurity collision integrals that allows us to relax the conserved charges to any equilibrium value we want. Our RTA is based on the realization that relaxation times pick out a preferred equilibrium that might differ from the local equilibrium used in the construction of our hydrodynamic model. Operationally, this amounts to restricting the possible values of the temperature and chemical potentials. Therefore, we use an RTA that picks out a reference background equilibrium. In addition, we allow for energy-dependent relaxations, which is the essential ingredient for obtaining generalized relaxation times. The relaxation times obtained from our modified RTA are Onsager reciprocal by default, upon fixing the hydrodynamic equilibrium to the one described by the reference Boltzmann distribution. 

 We begin with section \ref{sec:Magneto-transport} where we describe our hydrodynamic setup and show how it can be used to derive the fluid's thermoelectric conductivities. We then proceed in section \ref{sec:DC limit} to show how, at the level of hydrodynamics, different relaxation schemes constrain the relaxation times. In the same section, we introduce generalized relaxation times that are consistent with both Onsager reciprocity and phenomenological considerations. Finally, in sections \ref{sec:KinT} we show how our generalized relaxation times arise directly from kinetic theory.

\section{Hydrodynamic magneto-transport and the Onsager relations}
\label{sec:Magneto-transport}

In this section we briefly review the hydrodynamics of an anomalous fluid with two conserved charges: a $U(1)_V$ electric and a $U(1)_A$ axial charge, and we sketch out how to compute the fluid's longitudinal thermoelectric optical conductivities. We also discuss the constraints on transport arising from the Onsager relations. These are constraints due to imposing microscopic time-reversal invariance. 

An anomalous chiral fluid is characterized by the following (anomalous) conservation laws for energy-momentum, electric and chiral charge
\begin{subequations}\label{eqn:equations_of_motion}
\begin{align}
    \partial_\mu T^{\mu\nu}&=F^{\nu\lambda}J_\lambda~,\\
    \partial_\mu J^\mu&=0~,\\
    \partial_\mu J^\mu_5&=cE\cdot B~,
\end{align}
\end{subequations}
with $c$ being the anomaly coefficient, $T^{\mu\nu}$ the fluid's energy-momentum tensor and $J^\mu$ and $J^\mu_5$ the vector and chiral currents, respectively. The external electric $E^\mu$ and magnetic $B^\mu$ fields are defined in terms of the Maxwell field strength $F^{\mu\nu}$ and the fluid's four-velocity $u^\mu$ as $E^\mu = F^{\mu\nu}u_\nu$ and $B^\mu=\frac{1}{2}\epsilon^{\mu\nu\rho\sigma}u_{\nu}F_{\rho\sigma}$. To make use of \eqref{eqn:equations_of_motion}, we must write down the constitutive relations for $T^{\mu\nu}, J^\mu$ and $J^\mu_5$. To do so, we restrict ourselves to the case of an ideal fluid and a magnetic field $B$ which is ${\cal O}(1)$ in the hydrodynamic derivative expansion. This is the minimal and correct setup for discussing anomalous magneto-transport as shown in \cite{amoretti:framedependence}. The ideal constitutive relations for $T^{\mu\nu}, J^\mu$ and $J^\mu_5$ in a strong magnetic field are then given by\footnote{We have deliberately neglected some magnetization terms that are irrelevant for longitudinal transport to simplify the presentation.}
\begin{subequations}
    \begin{align}
    T^{\mu\nu}&=\epsilon u^\mu u^\nu + P\Delta^{\mu\nu}+\xi_\epsilon \left(B^\mu u^\nu+B^\nu u^\mu\right)~,\\
    J^\mu&=n u^\mu+\xi B^\mu~,\\
    J^\mu_5&=n_5 u^\mu+\xi_5 B^\mu~,
\end{align}
\end{subequations}
where $\Delta^{\mu\nu}=\eta^{\mu\nu}+u^\mu u^\nu$ is the projector orthogonal to the velocity and $P, \epsilon, n$ and $n_5$ are the fluid's pressure, energy density, electric charge density and axial charge density, respectively. The $\xi$ coefficients are sourced by the anomaly and read
\begin{equation}\label{eqn:anomalous_transport_coefficients}
    \xi=c\mu_5~~,~~ \xi_5=c\mu~~,~~ \xi_\epsilon=c\mu\mu_5~,
\end{equation}
where $\mu$ is the electric charge chemical potential and $\mu_{5}$ is the axial charge chemical potential. A non-zero axial chemical potential is seen in systems with asymmetrically filled Weyl cones. Such a situation does not render the system open, despite the resultant non-zero anomaly term, even though axial charge is not conserved. This should be compared to the relaxations we will introduce in the rest of the paper. Also, notice that there are no contributions from the mixed-gravitational anomaly in the magnetic field sector.

To derive the magneto-conductivities using hydrodynamics, we use linear response theory and consider the linearization of the equations of motion \eqref{eqn:equations_of_motion} around an equilibrium solution. In particular, we consider an equilibrium with constant temperature and chemical potentials, $T=\text{const}, \mu=\text{const}, \mu_5=\text{const}$. The fluid is considered at rest $u^\mu=(1,\vect{0})$ in the presence of a constant magnetic field $B^\nu = B \hat{z}$ and a vanishing electric field $E^\nu=0$. In terms of this background, the linearized equations of motion can be derived and solved to obtain the conductivities directly. Namely, we have
\begin{equation}\label{eqn:thermoelectric_matrix}
    \begin{pmatrix}
    \delta J_i\\
    \delta Q_i
    \end{pmatrix}=\begin{pmatrix}
    \sigma_{ij} &   T\alpha_{ij}\\
    T\Bar{\alpha}_{ij}  &   T\kappa_{ij}
    \end{pmatrix}
    \begin{pmatrix}
    \delta E_j\\
    \delta  \left(- \frac{\partial_j T}{T}\right)
    \end{pmatrix}~,
\end{equation}
where $\delta Q^i=\delta T^{0i}-\mu\delta J^i-\mu_5\delta J^i_5$ is the canonical heat current and the matrix on the right-hand side contains all the thermoelectric conductivities.

Apart from the conductivities, the linearized equations of motion also provide us with access to the Green's functions between the conserved charges. To be more precise, consider the linearized equations of motion for a two-charge fluid without anomalies. They can be written in a compact form via a spatial Fourier transform, 
\begin{equation}\label{eqn:schematic_equations_of_motion}
	\partial_t\varphi_a(t,\vect{k})+M_{ab}(\vect{k})\varphi_b(t,\vect{k})=0~,
\end{equation}
with $\varphi_a=\left(\delta\epsilon,\delta n,\delta n_5,\delta\pi^i\right)$  the fluctuations of the conserved charges. Here, $\pi^i$ is the fluid's spatial momentum while $M_{ab}$ is a matrix whose specific expression depends on the constitutive relations and the equations of motion. Subsequently, we can write down an analytic formula for the retarded Green's function $G^R_{ab}$ in linear response theory, namely \cite{kovtun:lecturenotes}
\begin{equation}\label{eqn:retarded_Green_function}
    G^R_{ab}(z, \mathbf{k}) = -(1 + izK^{-1}(z,\mathbf{k}))_{ac}\chi_{cb}~,
\end{equation}
where $z = \omega - i \eta$ with $\eta \rightarrow 0$, $K_{ab} = -iz\delta_{ab} + M_{ab}(\mathbf{k})$ and $\chi_{ab}=\frac{\partial\varphi_a}{\partial\lambda_b}$ is the thermodynamic susceptibility matrix\footnote{The $\lambda_a$ are given by $\lambda_a=\left(\frac{\delta T}{T},\delta\mu-\frac{\mu}{T}\delta T,\delta\mu_5-\frac{\mu_5}{T}\delta T,\delta v^i\right)$~.}
\begin{equation}
\label{eqn:SuscMatrix}
	\chi_{ab}=\begin{pmatrix}
        T\frac{\partial\epsilon}{\partial T}+\mu\frac{\partial\epsilon}{\partial\mu}+\mu_5\frac{\partial\epsilon}{\partial\mu_5}    &   \frac{\partial\epsilon}{\partial\mu}    &   \frac{\partial\epsilon}{\partial\mu_5}  &  0    &   0   &   0\\
        T\frac{\partial n}{\partial T}+\mu\frac{\partial n}{\partial\mu}+\mu_5\frac{\partial n}{\partial\mu_5}    &   \frac{\partial n}{\partial\mu}    &   \frac{\partial n}{\partial\mu_5}  &  0    &   0   &   0\\
        T\frac{\partial n_5}{\partial T}+\mu\frac{\partial n_5}{\partial\mu}+\mu_5\frac{\partial n_5}{\partial\mu_5}    &   \frac{\partial n_5}{\partial\mu}    &   \frac{\partial n_5}{\partial\mu_5}  &  0    &   0   &   0\\
        0   &   0   &   0   &   P+\epsilon  &   0   &   0\\
        0   &   0   &   0   &   0   &   P+\epsilon  &   0\\
        0   &   0   &   0   &   0   &   0   &   P+\epsilon
        \end{pmatrix}~.
\end{equation}
For later use, we also denote the $3\times3$ upper-left block of $\chi_{ab}$, whose entries are the energy and charge susceptibilities, by $\hat{\chi}$. 

For our purposes, the Green's function \eqref{eqn:retarded_Green_function} are useful because they allow us to enforce time-reversal covariance, i.e.~the Onsager relations, on the hydrodynamic transport coefficients. Namely, following \cite{kovtun:lecturenotes}, we require that 
\begin{equation}\label{eqn:Onsager_relations}
	G_{ab}^R(\omega,\vect{k};B )=\eta_a\eta_bG_{ba}^R(\omega ,-\vect{k};-B)~,
\end{equation}
where $\eta_a$ is the time-reversal eigenvalue for the field $\varphi_a$. This leads to the constraint\footnote{Recall the susceptibility matrix is symmetric, $\chi_{ab} = \chi_{ba}$.}
\begin{equation}\label{eqn:Onsager_hydrodynamics}
	\chi(B) SM^T(-\vect{k};-B)=M(\vect{k};B)\chi(B) S~,
\end{equation}
with $S=\text{diag}(1,1,1,-1,-1,-1)$ the matrix of time-reversal eigenvalues of the $\varphi$s. Note, that \eqref{eqn:Onsager_hydrodynamics} also takes into account that $B$ is odd under time-reversal.

One particular consequence of \eqref{eqn:Onsager_hydrodynamics} is the requirement that $\alpha=\Bar{\alpha}$ in the thermoelectric matrix defined in \eqref{eqn:thermoelectric_matrix}. Onsager relations have important consequences when we consider the DC limit of the thermoelectric conductivities, as we discuss in the following section.

\section{The DC limit and relaxation terms}
\label{sec:DC limit}

In the present section we want to focus on the DC, i.e.~zero frequency, limit of the magneto-conductivities derived via hydrodynamics. It is now well known that the anomalous conductivities become divergent in this limit \cite{landsteiner:negativemagnetoresistivity,lucas:hydrodynamictheory,abbasi:magnetotransportanomalous}, even when the magnetic field is treated as order zero in the derivative expansion \cite{amoretti:framedependence}. This divergence is a standard feature of non-anomalous hydrodynamics as well and arises due to the presence of an ever increasing total momentum generated by the external electric field. For non-anomalous hydrodynamics this issue can be resolved by simply adding momentum relaxation terms into the hydrodynamic equations (see for example \cite{Amoretti:2022acb,Amoretti:2022ovc} and citations therein).

Operationally, turning on momentum relaxation introduces one additional parameter into the system, the momentum-relaxation time $\tau_m$. For anomalous hydrodynamics, however, additional relaxation times besides $\tau_m$ are necessary. In particular, including axial charge, $\tau_5$, and energy relaxation, $\tau_\epsilon$, times into the equations of motion is also necessary \cite{landsteiner:negativemagnetoresistivity,lucas:hydrodynamictheory,abbasi:magnetotransportanomalous}. Where the need for these extra relaxations arises is clear from linearised hydrodynamics. As we mention above, in the presence of a constant electric field it is well understood that the Lorentz force, $\delta F^{i0}J_0=\delta E^in$, adds momentum without bound to the system, therefore momentum relaxation is necessary to avoid a blow-up. On the other hand, in standard non-anomalous hydrodynamics, the Joule heating $\delta(F^{0i}J_i)=\delta\vect{E}\cdot\delta\vect{J}\propto\delta\vect{E}\cdot\delta\vect{v}$ enters only at order two in fluctuations, hence energy relaxation is usually not necessary. However, in the anomalous case in the presence of a constant magnetic field $\vect{B}$, there is a non-zero current in the background ($\vect{J}=\xi\vect{B}$), so that the Joule heating appears at linear order in fluctuations $\delta(F^{0i}J_i)\propto\delta\vect{E}\cdot\vect{B}$. Thus, energy relaxation is also necessary in anomalous hydrodynamics to avoid infinite heating and obtain finite DC conductivity. Finally, axial charge relaxation is needed to balance the -- otherwise infinite -- axial charge injection caused by the anomaly itself $\vect{B}\cdot\delta\vect{E}$. The additional relaxations act like a soft cutoff for their respective conserved quantities, rendering the conductivities finite.

The existence of additional relaxation times besides $\tau_m$ can also be understood from a microscopic point of view in the case that anomalous hydrodynamics is realized in WSMs. In such systems, the axial charge relaxation time $\tau_5$ is always present, because the chiral symmetry is only approximate. In particular, the bands become non-linear beyond a particular value of the momentum, which renders the WSM quasiparticles massive and kills the anomaly. Furthermore when there are multiple Weyl nodes (required by topology in Weyl semimetals \cite{NIELSEN1981219,Nielsen:1980rz}), inter-valley scattering can deplete the axial charge. Energy relaxation is usually less relevant, but scatterings with phonons and inter-valley scattering can lead to energy relaxation. For completeness, we note that momentum relaxation $\tau_m$ is always present in condensed matter systems, due to the presence of impurities and phonons that take momentum away from the electron fluid.  

While the inclusion of finite $\tau_5,~\tau_m$ and $\tau_\epsilon$ are sufficient for obtaining finite conductivities, they are not necessary. It is possible to modify the relaxation terms by introducing additional, mixed relaxation times that couple e.g.~the energy and chiral charge equations of motion. In addition, we could keep only energy, momentum and chiral charge relaxation, but modify the charges which we relax (examples of this are given below). Both of these approaches modify the steady state and hence the short frequency response of the fluid. As a result, they generically give distinct DC conductivities. 

At the level of hydrodynamics, which approach we use is dictated by fundamental as well as phenomenological criteria. The fundamental criteria involve constraining the relaxation times such that the fluid is time-reversal invariant, satisfies the second law of thermodynamics\footnote{Violating the second law of thermodynamics is not such a grievous offense in general, as the relaxation terms can be thought of as rendering the fluid an open system.}, has finite DC conductivities and that electric charge is conserved. The phenomenological constraints take into account differences between relaxation times as observed in experiment, e.g.~typically $\tau_\epsilon \gg \tau_5$. In the subsections that follow we present several different realizations of relaxation terms in anomalous hydrodynamics and the DC transport they lead to. We show that only transport involving generalized relaxations can be made consistent with all fundamental and phenomenological constraints described above.

\subsection{Case 1: Canonical charge relaxation}
\label{sec:CanonChargeRel}

In the simplest case, used e.g.~in \cite{landsteiner:negativemagnetoresistivity,abbasi:magnetotransportanomalous,rogatko:magnetotransport}, we choose to relax the local charges as defined by their respective charge densities. For example, in order to relax the chiral charge, we can introduce a relaxation term into the equations of motion proportional to the chiral charge density as measured by the equilibrium observer, $\delta J^0_5$. The linearized equations of motion are subsequently modified to
\begin{subequations}\label{eqn:landsteiner_eom}
\begin{align}
    \partial_\mu\delta T^{\mu0}&=\delta(F^{0\lambda}J_\lambda)-\frac{1}{\tau_\epsilon}\delta T^{00}~,\\
    \partial_\mu\delta T^{\mu i}&=\delta(F^{i\lambda}J_\lambda)-\frac{1}{\tau_m}\delta T^{0i}~,\\
    \partial_\mu\delta J^\mu&=-\frac{1}{\tau_n} \delta J^{0}~, \label{eqn:chargerelaxation}\\
    \partial_\mu\delta J^\mu_5&=c\delta E\cdot B-\frac{1}{\tau_5}\delta J^0_5~.
\end{align}
\end{subequations}
While we have added an electric charge relaxation term, and a corresponding charge relaxation time $\tau_n$, to the right hand side of \eqref{eqn:chargerelaxation} this relaxation time does not appear in the AC conductivities and we can safely set it to zero.

The DC conductivities corresponding to \eqref{eqn:landsteiner_eom} have several interesting qualitative features. First of all, the DC thermoelectric conductivities contain terms generated by the anomaly. These terms induce a non-trivial magnetic field dependence, which differs from the usual $B^2$ dependence found in the literature. The $B^2$ dependence is, however, found in the limit that $B$ is ``small'' in amplitude. Furthermore, we confirm the results in \cite{abbasi:magnetotransportanomalous}, where a system with a single axial current is considered in the limit of vanishing chiral chemical potential ($\mu_5=n_5=\frac{\partial n_5}{\partial\mu}=\frac{\partial n_5}{\partial T}=0$). Namely, in the $\mu_5 =0$ limit the DC conductivities become
    \begin{subequations}
    \label{eq:ClassicConductivities}
    	\begin{align}
    		\sigma&=\frac{n^2\frac{\partial n_5}{\partial\mu_5}\tau_m+B^2c^2\left(sT\tau_5-n\mu\tau_m\right)}{\frac{\partial n_5}{\partial\mu_5}(P+\epsilon)-B^2c^2\mu^2}~,\\
    		\alpha&=\frac{s^2\frac{\partial n_5}{\partial\mu_5}T\tau_m}{\frac{\partial n_5}{\partial\mu_5}(P+\epsilon)-B^2c^2\mu^2}~,\\
    		\kappa&=\frac{s\left(n\frac{\partial n_5}{\partial\mu_5}-B^2c^2\mu\right)\tau_m}{\frac{\partial n_5}{\partial\mu_5}(P+\epsilon)-B^2c^2\mu^2}~.
    	\end{align}
    \end{subequations}
	In addition, \eqref{eq:ClassicConductivities} shows that energy relaxation is not necessary to obtain finite DC conductivities in the non-chiral regime, as the conductivities do not depend on $\tau_\epsilon$. From our above argument, this is expected as when $\mu_5=0$ there is no current in the background $\vect{J}=0$ and no Joule heating occurs.
	
Finally and most importantly, we note that for $\mu_{5} \neq 0$ one finds that $\alpha$ and $\kappa$ depend only on $\tau_m$, while $\sigma$ and $\bar\alpha$ depend also on $\tau_\epsilon$ and $\tau_5$. Thus enforcing Onsager reciprocity on the conductivities leads to constraints on the relaxation times. In particular, all relaxation times must be equal  $\tau_5=\tau_m=\tau_\epsilon=\tau_n$ \cite{abbasi:magnetotransportanomalous,amoretti:restoringtime}. This result is very difficult to justify phenomenologically as momentum relaxation is always present in metals and it should not be related to the axial charge recombination rate. Also energy and electric charge relaxation (if present at all) are usually much weaker effects. This disconnect with experiment suggests that we must modify the relaxation terms we included in the equations of motion. Such modifications are considered below.

\subsection{Case 2: Non-anomalous charge relaxation}\label{sec:relaxing_normal_charge}

A minimal modification of the relaxation scheme in \ref{sec:CanonChargeRel} is to relax only the normal charge densities, without including the anomaly-induced charges. This suggestion stems from the fact that the anomalous flow does not produce entropy, nor does it create drag or heat \cite{sadofyev:dragsuppression,stephanov:nodragframe}. For example, a chiral fluid in a constant magnetic field in the presence of impurities will relax the momentum such that the equilibrium velocity is zero. In contrast, the anomalous superfluid-like current will keep flowing without dissipation \cite{sadofyev:dragsuppression,stephanov:nodragframe}. Within this approach, the relaxation terms take the form
\begin{subequations}\label{eqn:normal_charge_relaxation}
\begin{align}
    \frac{1}{\tau_\epsilon}\delta T^{00}&=\frac{1}{\tau_\epsilon}\left(\delta\epsilon+2c\mu\mu_5\vect{B}\cdot\delta \vect{v}\right)&\longrightarrow& & &\frac{1}{\tau_\epsilon}\delta\epsilon~,\\
    \frac{1}{\tau_m}\delta T^{0i}&=\frac{1}{\tau_m}\left[(\epsilon+P)\delta v^i+B^ic(\mu_5\delta\mu+\mu\delta\mu_5)\right]&\longrightarrow& & &\frac{1}{\tau_m}(\epsilon+P)\delta v^i~,\\
    \frac{1}{\tau_n}\delta J^0&=\frac{1}{\tau_n}\left(\delta n+c\mu_5\vect{B}\cdot\delta\vect{v}\right)&\longrightarrow& & &\frac{1}{\tau_n}\delta n~,\\
    \frac{1}{\tau_5}\delta J^0_5&=\frac{1}{\tau_5}\left(\delta n_5+c\mu\vect{B}\cdot\delta\vect{v}\right)&\longrightarrow& & &\frac{1}{\tau_5}\delta n_5~.
\end{align}
\end{subequations}
This partially resolves the issue of equal relaxation times from the previous section. In particular, the momentum relaxation time $\tau_m$ now decouples from the rest and is not constrained by time-reversal invariance. However, Onsager relations \eqref{eqn:Onsager_hydrodynamics} still impose that energy, axial and electric charge all relax at the same rate $\tau_n=\tau_5=\tau_\epsilon$ (this time $\tau_n$ appears in the conductivities). Again this is hard to justify from a microscopic picture, not only because energy and chiral charge are expected to relax at different rates, but also because we expect electric charge to be conserved identically. In other words, after all constraints have been imposed, we would like to be able to set $\tau_n^{-1} = 0$ without issue. 

Finally, as in the previous subsection, we can arrive at finite DC conductivities without including energy relaxation in the non-chiral limit, $\mu_5=n_5=\frac{\partial n_5}{\partial\mu}=\frac{\partial n_5}{\partial T}=0$, where
	\begin{equation}\label{eqn:conductivity_zero_axial}
		\sigma=\frac{n^2\tau_m}{P+\epsilon}+\frac{B^2c^2\tau_5}{\frac{\partial n_5}{\partial\mu_5}}~.
	\end{equation}
 This is where the similarities to the previous case end however. In general the AC conductivities due to the non-anomalous charge relaxation terms \eqref{eqn:normal_charge_relaxation} depend on all four relaxation times. Furthermore, in the DC limit they all take the non-anomalous hydrodynamic form\footnote{\label{Footnote} Although $\kappa$ is not anomalous in DC, the thermal conductivity measured in experiments is related to this one by $\bar\kappa=\kappa-T\alpha^2/\sigma$, thus $\bar\kappa$ is anomalous in DC, due to $\sigma$ being anomalous.}, except for the electric conductivity which remains anomalous even at $\omega=0$ with a simple $B^2$ dependence. This can be understood from two related observations: first, the linearized heat current is not anomalous (i.e.~$\delta Q^i=sT\delta v^i$), and second, the anomalous non-dissipative transport coefficients \eqref{eqn:anomalous_transport_coefficients} do not have any $T$-dependence induced by the mixed-gravitational anomaly.
 
We see that relaxing the normal charge has had some interesting consequences. For example, in other theoretical studies where the DC conductivity is explicitly reported \cite{lucas:hydrodynamictheory,chernodub:thermaltransport,PhysRevResearch.2.013088,PhysRevB.93.085107,sharma:nernst_magnetothermal,lundgren:thermoelectric_properties}, the anomaly appears in the thermo-electric conductivity from hydrodynamic, lattice and kinetic theory approaches; whether it appears in our thermal conductivity $\kappa$ depends on definition (see footnote \ref{Footnote}). To our knowledge this is because the authors relax the total charge, and use differing assumptions such as having two distinct fluid velocities (see e.g.~\cite{lucas:hydrodynamictheory}). In principle this difference is open to experimental verification \cite{Vu:thermal_chiral_anomaly,Gooth_2017,Jia2016ThermoelectricSO}. However, a difficulty in doing so is that the one of the accepted signatures of the anomaly, NMR, can be generated by other mechanisms.

The general message to take away from this section is that we can decouple momentum relaxation from all other relaxing charges by neglecting the anomalous contributions to said charges. This, however, forces charge relaxation into the equations in order to generate Onsager reciprocal conductivities. These results in conjunction with those of section \ref{sec:CanonChargeRel} suggest that we must consider a more exotic version of charge relaxation in order to conform to both fundamental and phenomenological constraints. We do so in the following section.

\subsection{Case 3: Generalized relaxations}
\label{sec:generalised relaxations}

The previous two sections, \ref{sec:CanonChargeRel} and \ref{sec:relaxing_normal_charge}, represent work either already in the literature, or a slight modification thereof, and have the problem that normal charge relaxes if we desire a finite DC conductivity. Our second suggestion (c.f. section \ref{sec:relaxing_normal_charge}) for choosing appropriate relaxation terms follows from the idea of generalized relaxations \cite{lucas:hydrodynamictheory,amoretti:restoringtime}. Within this approach, one adds mixed relaxation times that couple the various relaxing charges. These additional relaxation terms provide us with enough parameters to allow for finite and Onsager reciprocal DC conductivities, without requiring electric charge relaxation.

In keeping with the results of the previous section, we consider generalized relaxations only between the energy, charge and axial charge equations of motion and include the usual momentum relaxation time $\tau_m$ in the momentum equation. Thus, consider the generic relaxations that modify the RHS of \eqref{eqn:landsteiner_eom} to
\begin{equation}\label{eqn:generic_relaxations}
    \begin{rcases*}
        \text{energy:}\hspace{1.3cm} \frac{1}{\tau_{\epsilon\epsilon}}\delta\epsilon+\frac{1}{\tau_{\epsilon n}}\delta n+\frac{1}{\tau_{\epsilon n_5}}\delta n_5\\
        \text{charge:}\hspace{1.3cm}\frac{1}{\tau_{n\epsilon}}\delta\epsilon+\frac{1}{\tau_{nn}}\delta n+\frac{1}{\tau_{nn_5}}\delta n_5\\
        \text{axial charge:}\quad\frac{1}{\tau_{n_5\epsilon}}\delta\epsilon+\frac{1}{\tau_{n_5n}}\delta n+\frac{1}{\tau_{n_5n_5}}\delta n_5
    \end{rcases*} =\hat{\tau}\cdot\varphi~,
\end{equation}
where $\varphi = (\delta\epsilon, \delta n, \delta n_5)$ and $\hat{\tau}$ is the $3\times 3$ relaxation time matrix. The mixed relaxation time $\tau_{\epsilon n}$ represents the energy relaxation due to charge fluctuations and similarly for the other mixed relaxation terms. Notice that the mixed relaxations involving energy and charge do not have units of time and they are not in general expected to have a definite sign.

As always, Onsager reciprocity constrains the relaxation times. Namely, $\hat{\tau}$ must satisfy

\begin{subequations}\label{eqn:Onsager_constraints}
\begin{align}
    0&=\frac{\chi_{nn_5}}{\tau_{\epsilon n_5}}+\frac{\chi_{nn}}{\tau_{\epsilon n}}-\frac{\chi_{\epsilon n_5}}{\tau_{nn_5}}+\frac{\chi_{\epsilon n}}{\tau_{\epsilon\epsilon}}-\frac{\chi_{\epsilon n}}{\tau_{nn}}-\frac{\chi_{\epsilon\epsilon}}{\tau_{n\epsilon}}~,\\
    0&=\frac{\chi_{n_5n_5}}{\tau_{\epsilon n_5}}+\frac{\chi_{nn_5}}{\tau_{\epsilon n}}-\frac{\chi_{\epsilon n_5}}{\tau_{n_5n_5}}+\frac{\chi_{\epsilon n_5}}{\tau_{\epsilon\epsilon}}-\frac{\chi_{\epsilon n}}{\tau_{n_5n}}-\frac{\chi_{\epsilon\epsilon}}{\tau_{n_5\epsilon}}~,\\
    0&=\frac{\chi_{n_5n_5}}{\tau_{nn_5}}-\frac{\chi_{nn_5}}{\tau_{n_5 n_5}}+\frac{\chi_{n n_5}}{\tau_{nn}}-\frac{\chi_{nn}}{\tau_{n_5n}}+\frac{\chi_{\epsilon n_5}}{\tau_{n\epsilon}}-\frac{\chi_{\epsilon n}}{\tau_{n_5\epsilon}}~,
\end{align}    
\end{subequations}
or in matrix formulation
\begin{equation}
\label{Eq:XTCommutator}
    \hat\tau\cdot\hat\chi = \left(\hat\tau\cdot\hat\chi\right)^T ~,
\end{equation}
where $T$ denotes the matrix transpose and $\hat{{\chi}}$ is defined around equation \eqref{eqn:SuscMatrix}.

Our goal now is to study the parameter space of these generalized relaxations to find regions that allow for finite and Onsager reciprocal DC conductivities. Consider first the case of a conserved electric charge, but relaxed energy, momentum and axial charge. In particular, consider the minimal set of relaxations satisfying \eqref{Eq:XTCommutator} after setting $\tau_{nn}^{-1}=\tau_{n\epsilon}^{-1}=\tau_{nn_5}^{-1}=0$. Thus, starting from 9 relaxations, we are left with 6 non-zero relaxation parameters and 3 equations \eqref{eqn:Onsager_constraints}. Only 3 parameters are independent, so we may assume $\tau_{\epsilon\epsilon}$, $\tau_{n_5n_5}$ and $\tau_{n_5 \epsilon}$ are completely independent and express the remaining relaxation times in terms of them. Interestingly, it is also possible to satisfy \eqref{eqn:Onsager_constraints} after setting either $\tau_{\epsilon n}^{-1}=0$ or $\tau_{n_5n}^{-1}=0$ leaving us with two independent parameters. Either choice leads by default to Onsager reciprocal conductivities, however finiteness in the DC limit is not guaranteed. Computing the thermoelectric conductivities, if only one of $\tau_{\epsilon n}^{-1}$ or $\tau_{n_5n}^{-1}$ is non-zero then we find finite DC conductivities. However, if both are simultaneously zero then the DC conductivities cease to be finite.

On top of this Onsager reciprocal solution, we can further try to satisfy the second law of thermodynamics. The second law imposes three additional constraints on the relaxation times \cite{amoretti:restoringtime}
\begin{subequations}
\begin{align}
    0&=\frac{1}{\tau_{\epsilon\epsilon}}-\frac{\mu}{\tau_{n\epsilon}}-\frac{\mu_5}{\tau_{n_5\epsilon}}~,\\
    0&=\frac{1}{\tau_{\epsilon n_5}}-\frac{\mu}{\tau_{nn_5}}-\frac{\mu_5}{\tau_{n_5n_5}}~,\\
    0&=\frac{1}{\tau_{\epsilon n}}-\frac{\mu}{\tau_{nn}}-\frac{\mu_5}{\tau_{n_5n}}~.
\end{align}
\end{subequations}
Hence we end up with 6 parameters and 6 equations, whose only solution is the trivial one. Thus, with the structure we have here, we observe that it is not possible to have time-reversal covariance, electric charge conservation and satisfy the second law of thermodynamics even if generalized relaxations are considered. Put differently, a fluid consisting of one conserved normal charge and one axial charge which satisfies Onsager reciprocity is always open.

Finally, we mention in passing that we can also have a minimal setup for the relaxation times, that leads to Onsager reciprocal and DC-finite conductivities, if we allow for weak electric charge relaxation (which if present must be parametrically small). In this case we can, for example, impose $\tau_{\epsilon n}^{-1}=\tau_{n\epsilon}^{-1}=\tau_{n_5n}^{-1}=\tau_{nn_5}^{-1}=0$. Subsequently, we again have five non-zero parameters and three constraints from the Onsager relations. This leaves us with two independent relaxation rates, which we can take to be $\tau_{n_5n_5}$ and $\tau_{nn}$. We have checked explicitly that the DC conductivities derived within this setup are finite.

\section{Kinetic theory of generalized relaxations}
\label{sec:KinT}

In the present section, we show that generalized relaxations satisfying Onsager reciprocity can arise via microscopic considerations and particularly a modification of the relaxation time approximation (RTA) in kinetic theory. Essentially, our RTA employs a reference Boltzmann distribution function to which we relax the electron distribution function via \textit{energy-dependent} relaxation times. The reference distribution function is chosen  consistent with the hydrodynamic equilibrium we use in section \ref{sec:generalised relaxations}. This approach is based on \cite{gorbar:consistenthydrodynamics,narozhny:hydrodynamicapproach}, where it was solely used to introduce momentum relaxation.

Notice that the relations between the generalized relaxations are independent of the anomaly; as soon as we relax only the normal charge, we decouple the anomaly from the relaxation rates. This means that from a kinetic theory perspective we can use non-chiral kinetic theory. For this reason we first elucidate our novel RTA on a fluid conserving energy, momentum and electric charge, which interacts with impurities and phonons. We then proceed to a two-current model (representing the vector and chiral currents in a Weyl semimetal) in section \ref{sec:twocurrents} to discuss the case of interest.

We will proceed by first postulating a reasonable RTA for our system that leads to the desired relaxations, and subsequently justify it with a more refined microscopic collision integral.

\subsection{Single current model}
\label{sec:SingleCurrent}

Consider first the Boltzmann equation for the one-particle distribution function $f(\vect{x},\vect{p},t)\equiv f_{\vect{p}}$ of massless relativistic particles. In the absence of external forces it reads\footnote{External forces are responsible for the Maxwell term in the hydrodynamic limit and are not relevant when discussing generalized relaxations. }
\begin{equation}
\label{eq:BoltzmannEq}
    \partial_tf_{\vect{p}}+\vect{p}\cdot\vect{\nabla}f_{\vect{p}}=I_\text{coll}[f_{\vect{p}}]~.
\end{equation}
$I_\text{coll}$ is the collision integral and is a functional of $f_{\vect{p}}$. Typically \cite{huang:statisticalmechanics, tong:statisticalphysics, denicol:microscopicfoundations,landau:physicalkinetics}, $I_\text{coll}$ for electron-electron scattering is assumed to conserve energy, momentum and charge, i.e. 
\begin{equation}
    \int\frac{\dif^3\vect{p}}{(2\pi)^3} A(\vect{x},\vect{p})I_\text{ee}[f_{\vect{p}}]=0\qquad\qquad\text{for}\quad A=\{\epsilon_p,\vect{p},1\}~,
\end{equation}
where $\epsilon_p$ is the single-particle energy. In addition, it is assumed $I_\text{ee}$ respects unitarity, time-reversal, parity symmetry and molecular chaos (i.e.~the distributions of the incoming particles are uncorrelated). Under these assumptions the electron-electron collision integral, vanishes identically when $f_{\vect{p}}$ is given by the distribution of \textit{local} thermal equilibrium (LTE) 
\begin{equation}\label{eqn:LTE_solution}
    f^\text{eq}_{\vect{p}}=\frac{1}{1 + e^{(\epsilon_p-\vect{u}\cdot\vect{p}-\mu)/T}}~,
\end{equation}
where the Lagrange multipliers $T$, $\vect{u}$ and $\mu$ are functions of space (but not of momentum). The equations of ideal hydrodynamics follow by multiplying the full Boltzmann equation with energy, momentum or the identity and integrating over momentum after substituting $f_{\vect{p}}\rightarrow f^{\rm eq}_{\vect{p}}$. In particular the RHS of \eqref{eq:BoltzmannEq} is zero, hence energy, momentum and charge are identically conserved.

To introduce relaxation, specifically momentum relaxation, we follow the approach in \cite{gorbar:consistenthydrodynamics,narozhny:hydrodynamicapproach}. We start from the full LTE solution $f^{\rm eq}_{\vect{p}}$---which we denote henceforth by $f_{\vect{p}}$ to simplify the notation---and expand it at small velocity as 
\begin{equation}
    f_{\vect{p}}\approx f^{(0)}+(\vect{p}\cdot\vect{u})\frac{\partial f^{(0)}}{\partial\epsilon_p}\qquad\text{with}\qquad f^{(0)}=\frac{1}{1+e^{(\epsilon_p-\mu)/T}}~.
\end{equation}
Because of the linearization at small velocity, only the $f^{(0)}$ term actually contributes to the equilibrium energy and charge, as the velocity dependent term drops out when integrated in momentum space \cite{gorbar:consistenthydrodynamics}. For the case of WSMs, the energy dispersion relation is $\epsilon_p=p$ and the charges are
\begin{subequations}
	\begin{align}
		\epsilon&=\frac{15\mu^4+30\pi^2T^2\mu^2+7\pi^4T^4}{120\pi^2}~,\\
		n&=\mu\left(\frac{\mu^2}{6\pi^2}+\frac{T^2}{6}\right)~,
	\end{align}    
\end{subequations}
where energy and charge density are measured from the Weyl point.

We can now introduce the appropriate RTA ansatz that relaxes momentum \cite{gorbar:consistenthydrodynamics,narozhny:hydrodynamicapproach}. In general, within RTA we assume that the collision integral takes a particular simple form that depends on the 1-particle distribution $f_{\vect{p}}$ entering the Boltzmann equation \eqref{eq:BoltzmannEq} and a reference 1-particle distribution to which $f^{\rm ref}_{\vect{p}}$ relaxes to. In our case, we assume that
\begin{equation}
\label{eqn:momentum_relaxation}
    I_\text{ei}\approx-\frac{f_{\vect{p}}-f^{(0)}}{\tau_m}~.
\end{equation}
The above RTA is quite different from the standard RTA in kinetic theory, in the sense that both $f_{\vect{p}}$ and $f^{(0)}$ in \eqref{eqn:momentum_relaxation} are LTE solutions for the ideal fluid, i.e.~they make the standard electron-electron integral vanish by default. The physical justification for such a term is the following: momentum relaxation behaves as an extra constraint on our theory and is such that, of all the possible equilibrium solutions at constant velocity (related by boosts), only one equilibrium is actually picked by the system - the one at zero velocity identified with $f^{(0)}$. Hence while all LTE solutions are valid relaxation targets in principle (they all respect detailed balance), we work with an ansatz for the RTA that selects only the zero velocity LTE as the solution for which $I_{\rm coll} = 0$.  This fact is reflected in the hydrodynamic equation for momentum which now includes a momentum relaxation term\footnote{Energy and charge are identically conserved.}
\begin{equation}
    \partial_t(\epsilon+P)v^i+\dots=-\frac{(\epsilon+P)v^i}{\tau_m}~.
\end{equation}
An RTA of the form \eqref{eqn:momentum_relaxation} can be derived by solving the detailed balance equation for certain electron-impurity elastic collision integrals, see e.g.~\cite{landau:physicalkinetics,pal:resistivity}. In our case we can think of the collision integral taking the form \cite{landau:physicalkinetics}
\begin{equation}
\label{eq:EEandEI_Collision}
    I_\text{coll}=I_\text{ee}+I_\text{ei}~,
\end{equation}
where $I_{\rm ee}$ is the standard electron-electron collision integral forcing LTE as a solution of the Boltzmann equation, while $I_\text{ei}$ describes elastic electron-impurity scatterings and takes the form
\begin{equation}\label{eqn:impurities_collision_integral}
    I_\text{ei}=\int\dif^3\vect{p}'W_{\vect{p}\rightarrow\vect{p}'}[f_{\vect{p}}-f_{\vect{p}'}]\delta(\epsilon_p-\epsilon_{p'})~,
\end{equation}
with $W_{\vect{p}\rightarrow\vect{p}'}$ the electron-impurity scattering rate from electron momentum $\vect{p}$ to $\vect{p}'$. Following the usual approach, we look for distributions (among the spectrum of all LTE solutions) that make $I_{\rm ei}$ vanish. These are indeed the zero velocity distributions $f^{(0)}$. Then, assuming $I_{\rm ei}$ takes an RTA form, we are led to \eqref{eqn:momentum_relaxation}.

The above suggests we can add energy and charge relaxation to hydrodynamics by including an additional collision integral into $I_{\rm coll}$ that depends on a 1-particle distribution $\bar{f}^{(0)}$ that picks an LTE with fixed energy and charge $\bar{\epsilon}$ and $\bar{n}$ respectively. This is indeed possible by using an RTA of the form\footnote{We note that a similar collision integral has been suggested in \cite{DASH2022137202}. The difference between our collision integral and theirs is that their collision integral is first order in derivatives.}
\begin{equation}\label{eqn:charge_relaxation}
    I_\text{coll}=I_\text{ee}-\frac{f_{\vect{p}}-f^{(0)}}{\tau'_m}-\frac{f_{\vect{p}}-\Bar{f}^{(0)}}{\tau_n}=I_\text{ee}-\frac{f_{\vect{p}}-f^{(0)}}{\tau_m}-\frac{f^{(0)}-\bar{f}^{(0)}}{\tau_n}~,
\end{equation}
where 
\begin{equation}
    \Bar{f}^{(0)}=\frac{1}{1+e^{(\epsilon_p-\bar\mu)/\bar T}}
\end{equation}
depends on the reference values of temperature $\Bar{T}$ and chemical potential $\Bar{\mu}$. In the second equality in \eqref{eqn:charge_relaxation}, we redefined $\tau_m$ in order to separate the relaxation terms with different properties: The momentum relaxation term is zero on the energy and charge equation and only relaxes momentum as desired. The second term instead vanishes in the momentum equation, and relaxes energy and charge at the same rate $\tau_n$
\begin{subequations}
\begin{align}
    \partial_t\epsilon+\dots&=-\frac{\epsilon-\Bar{\epsilon}}{\tau_n}~,\\
    \partial_t(P+\epsilon)v^i+\dots&=-\frac{(P+\epsilon)v^i}{\tau_m}~,\\
    \partial_tn+\dots&=-\frac{n-\bar{n}}{\tau_n}~.
\end{align}
\end{subequations}
Thus, we see that the relaxed hydrodynamic equations discussed in \ref{sec:relaxing_normal_charge} follows directly from an RTA of the form \eqref{eqn:charge_relaxation}. From now on we denote with a bar thermodynamic quantities computed with respect to $\bar{f}^{(0)}$.

Similarly to the momentum relaxation term, the energy and charge relaxation terms follow from microscopic considerations as well. Namely, the corresponding collision integral for the electrons is schematically written as
\begin{equation}
    I_\text{coll}=I_\text{ee}+I_\text{ei}+I_\text{ep}~,
\end{equation}
where the role of $I_{\rm ee}$ and $I_{\rm ei}$ were discussed around \eqref{eq:EEandEI_Collision} while the electron-phonon collision integral takes the following form \cite{landau:physicalkinetics}
\begin{align}
	I_\text{ep}&=\int\dif^3\vect{q}W_{\vect{p}',\vect{q}\rightarrow\vect{p}}\left[f_{\vect{p}'}(1-f_{\vect{p}})n_{\vect{q}}-f_{\vect{p}}(1-f_{\vect{p}'})(1+n_{\vect{q}})\right]\delta\left(\epsilon_p-\epsilon_{p'}-\omega_q\right)+ \nonumber \\
	&\quad+\int\dif^3\vect{q}W_{\vect{p}'\rightarrow\vect{p},\vect{q}}\left[f_{\vect{p}'}(1-f_{\vect{p}})(1+n_{\vect{q}})-f_{\vect{p}}(1-f_{\vect{p}'})n_{\vect{q}}\right]\delta\left(\epsilon_p+\omega_k-\epsilon_{p'}\right) \; .
\end{align}
Here, $\omega_q$ and $n_{\vect{q}}$ are respectively the energy and distribution function for the phonon, while $W$ is the effective scattering vertex. The first term represents processes in which a quasi-electron with momentum $\vect{p}$ emits a phonon with momentum $\vect{q}$ and also the reverse process where an electron with momentum $\vect{p}'$ absorbs a phonon of momentum $\vect{q}$. The second term corresponds to absorption of a phonon by an electron with momentum $\vect{p}$ and the reverse process of emission.  Importantly, the sum of both contributions (absorption and emission from higher or lower energy states) conserves the total electric charge. However, $I_\text{ep}$ vanishes only when $f_{\vect{p}}$ is in global thermal equilibrium. This thermal equilibrium can be fixed by assuming, as in \cite{narozhny:energyrelaxation}, that the phonons are in thermal equilibrium and act as a bath for the electrons.

If we now want to write $I_{\rm ep}$ in an RTA form, we should subtract from $f_{\vect{p}}$ the global distribution $\Bar{f}^{(0)}$, so that the RTA takes exactly the form \eqref{eqn:charge_relaxation} and vanishes when $\Bar{\mu}$ and $\Bar{T}$ are in global equilibrium. This way we learn that $\Bar{\mu}$ and $\Bar{T}$ do not describe a generic LTE distribution function, but rather a global one.

One comment here is in order, the electron-phonon collision integrals presented above  conserve the total electric charge because they vanish when integrated in momentum space. This, however, is not true for their RTA form \eqref{eqn:charge_relaxation} which relaxes both energy and charge. This is not an issue since it is well known that the RTA does not inherit all the properties of the true collision integral \cite{cercignani:mathematicalmethods}. The only property the RTA must have is that it vanishes when the distribution function takes on values that set to zero the true collision integral. From this point of view, charge relaxation is a feature specific to the RTA and not a property inherited from a microscopic picture.\footnote{One could obviously also consider scatterings with charged impurities that relax the total charge from a microscopic point of view.} This is in agreement with the results of section \ref{sec:relaxing_normal_charge} where time-reversal invariance forbids setting $\tau_n^{-1}$ to zero and forces us to introduce generalized relaxations into the hydrodynamic equations. 

Now, we want to generalize our RTA to include generalized relaxations. As in the hydrodynamics section \ref{sec:generalised relaxations}, we look for generalized relaxations between energy and charge, but leave momentum relaxation decoupled from all the other charges. This means that we need to modify only the last term in the collision integral \eqref{eqn:charge_relaxation}. To motivate our RTA ansatz, we note that the RTA is a very crude approximation for the collision integral in which all the scattering dynamics is expressed in terms of a single constant parameter. In general however, the scattering rate will depend on the energy and consequently a minimal modification for our purposes is to assume that the relaxation time $\tau_{n}$ depends on energy \cite{pal:resistivity}. Such a modification is expected for electrons scattering with impurities and phonons, see e.g.~\cite{ochi:electronhole}.

To be precise our RTA ansatz follows by expanding the energy-dependent relaxation time $\tau_{n}(\epsilon_p)$ for the charge, such that the collision integral takes the power-series form (modulo momentum relaxing terms)
\begin{equation}\label{eqn:generalised_relaxations_collision_integral}
    I_\text{coll}=\sum_{j \geq-2}\epsilon_p^j\frac{f^{(0)}-\Bar{f}^{(0)}}{\tau_{j}}=\frac{1}{\epsilon_p^2}\frac{f^{(0)}-\Bar{f}^{(0)}}{\tau_{-2}}+\frac{1}{\epsilon_p}\frac{f^{(0)}-\Bar{f}^{(0)}}{\tau_{-1}}+\frac{f^{(0)}-\bar{f}^{(0)}}{\tau_0}+\dots
\end{equation}
The terms with negative powers of $\epsilon_p$ are the only ones that lead to finite quantities when integrated in momentum space for a linear dispersion relation $\epsilon_p=p$. More precisely, \eqref{eqn:generalised_relaxations_collision_integral} leads to the following hydrodynamic equations
\begin{subequations}
\label{eq:1currentGenRel}
    \begin{align}
    \partial_t\epsilon+\dots&=-\frac{K-\bar K}{\tau_{-2}}-\frac{n-\bar n}{\tau_{-1}}-\frac{\epsilon-\bar\epsilon}{\tau_0}-\frac{L-\bar L}{\tau_1}+\dots~,\\
    \partial_t n+\dots&=-\frac{A-\bar A}{\tau_{-2}}-\frac{K-\bar K}{\tau_{-1}}-\frac{n-\bar n}{\tau_0}-\frac{\epsilon-\bar\epsilon}{\tau_1}+\dots~,
\end{align}
\end{subequations}
where $K,L$ and $A$ (and also the dots) represent thermodynamic functions that can be analytically computed given a specific power of $\epsilon_p$ in the collision integral. For example, the first few terms for massless relativistic fermions are
\begin{subequations}
    \begin{align}
    A&=\frac{\mu}{2\pi^2}~,\\
    K&=\frac{T^2}{4\pi^2}\left(\frac{\mu^2}{T^2}+\frac{\pi^2}{3}\right)~,\\
    L&=\frac{T^4\mu}{30\pi^2}\left(\pi^2+\frac{\mu^2}{T^2}\right)\left(7\pi^2+3\frac{\mu^2}{T^2}\right)~.
\end{align}
\end{subequations}

As already discussed in \cite{amoretti:restoringtime}, the linearized hydrodynamic theory is agnostic on the full non-linear form of the hydrodynamic equations. For this reason, in order to match the generalized relaxations in \eqref{eq:1currentGenRel} with those in section \ref{sec:generalised relaxations}, we must linearize \eqref{eq:1currentGenRel} around a fixed background. We linearise around the reference values $\bar\epsilon$ and $\bar n$ with fluctuations $\delta\epsilon$ and $\delta n$ to obtain 
\begin{subequations}
    \begin{align}
    \partial_t\delta\epsilon+\dots&=-\frac{\delta\epsilon}{\tau_{\epsilon\epsilon}}-\frac{\delta n}{\tau_{\epsilon n}} \; , \\
    \partial_t \delta n+\dots &=-\frac{\delta\epsilon}{\tau_{n\epsilon}}-\frac{\delta n}{\tau_{nn}} \; , 
\end{align}
\end{subequations}
where we defined the usual effective relaxation rates in terms of the microscopic $\tau_{j}$ that appeared in the kinetic theory collision integral, e.g.
\begin{align}
    \frac{1}{\tau_{nn}}=\frac{\partial A}{\partial n}\frac{1}{\tau_{-2}}+\frac{\partial K}{\partial n}\frac{1}{\tau_{-1}}+\frac{1}{\tau_0}+\dots \; . 
\end{align}
The dots represent the contribution of terms in the collision integral with $\tau_{j}$, $j\geq2$.  We observe that this final expression is reminiscent of Matthiessen's rule. 
Having identified the generalized relaxation times, we can now confirm that they respect Onsager reciprocity. In particular, they identically satisfy
\begin{equation}\label{eqn:one_current_Onsager}
    \frac{\chi_{\epsilon\epsilon}}{\tau_{n\epsilon}}-\frac{\chi_{\epsilon n}}{\tau_{\epsilon\epsilon}}+\frac{\chi_{n\epsilon}}{\tau_{nn}}-\frac{\chi_{nn}}{\tau_{\epsilon n}}=0~.
\end{equation}
Note that as we found in section \ref{sec:generalised relaxations}, the second law of thermodynamics is not necessarily identically satisfied. From a physical perspective this should be expected since our system interacts with phonons and impurities and is therefore weakly open.

All of the above confirms that our generalized hydrodynamic relaxations can arise from a microscopic viewpoint and particularly kinetic theory, if we assume the generalized RTA \eqref{eqn:generalised_relaxations_collision_integral} for the collision integral. While we have proven this result for a relativistic dispersion relation we expect it to hold for a generic dispersion relation $\epsilon_p$ and for scattering times $\tau$ which are suitable non-analytic functions of $\epsilon_p$.

Before we proceed to the case of WSMs and two conserved currents, some comments are in order. First, note that if instead of keeping the full series in \eqref{eqn:generalised_relaxations_collision_integral} we had truncated it to the first couple of terms, then we would still have found that the Onsager relations are obeyed. However, in this case the four, a priori different, $\tau$ of the generalized relaxations would be functions of only one or two $\tau_{j}$. Hence, the generalised relaxations could actually be expressed in terms of each other. These expressions cannot be detected by means of hydrodynamics alone, and a fully microscopic approach is necessary in order to derive them.
 
Second, one might wonder what happens if we include energy corrections to the momentum relaxation collision integral \eqref{eqn:momentum_relaxation}. In this case, one finds new functions of the thermodynamics multiplying the velocity. However when we linearise these new relaxations can always be written as $(P+\epsilon)\delta\vect{v}/\tau_m$ by redefining $\tau_m$ appropriately. Hence, energy corrections do not modify the linearized momentum (non-)conservation equation, they only change the value of $\tau_m$.

Finally, we can ask ourselves whether we can enforce charge conservation within this setup. In particular, can we impose
\begin{equation}
    \int\frac{\dif^3p}{(2\pi)^3}\frac{f^{(0)}-\bar{f}^{(0)}}{\tau(\epsilon_p)}=\sum_{j\geq-2}\int\frac{\dif^3p}{(2\pi)^3}\epsilon_p^j\frac{f^{(0)}-\bar{f}^{(0)}}{\tau_{j}}=0~?
\end{equation}
This is not always possible within the RTA presented in this section. For example, if the only relevant terms in the $\tau_{n}(\epsilon_p)$ expansion are $\tau_0$ and $\tau_1$, then $\tau_{nn}=\tau_0$, $\tau_{n\epsilon}=\tau_1$ and setting charge relaxation to zero amounts to $\tau_0^{-1}=0=\tau_1^{-1}$. However by including additional terms in the expansion, we can in principle impose charge conservation by requiring that the microscopic $\tau_{j}$ obey a certain set of constraints. It is not obvious that these constraints necessarily follow directly from the microscopics and might require some fine-tuning.

\subsection{Two current model}
\label{sec:twocurrents}

We can now generalize our approach to the case of a system with one single conserved stress energy tensor, but two conserved currents. In this case, we need to take into account two different distribution functions $f_{\vect{p},\lambda}$, with $\lambda=\pm$ denoting the chirality of each species of electron located around its corresponding Weyl node  \cite{son:kinetictheory,gorbar:consistentchiralkinetic,dantas:magnetotransportmultiweyl}. The most general form of the Boltzmann equation satisfied by $f_{\vect{p},\lambda}$ is \cite{pongsangangan:hydrodynamics,bianca:boltzmanngasmixture,fotakis:multicomponent,landau:physicalkinetics}
\begin{equation}
\label{eq:ChiralBoltz}
    \partial_t \vec{f}_{\vect{p}}+\vect{p}\cdot\vect{\nabla}\vec{f}_{\vect{p}}=I_{ee}[\vec{f}_{\vect{p}}]=\begin{pmatrix}
        I_{++}[f_{\vect{p},+}]  +   I_{+-}[f_{\vect{p},\lambda}]\\
        I_{-+}[f_{\vect{p},\lambda}]  +   I_{--}[f_{\vect{p},-}]
    \end{pmatrix}~,
\end{equation}
where $\vec{f}_{\vect{p}}=(f_{\vect{p},+},f_{\vect{p},-})^T$. The mixing terms of the electron-electron collision integral on the right hand side of \eqref{eq:ChiralBoltz} imply that fermions with opposite chiralities interact with each other and hence thermalize simultaneously, whenever $I_{\pm\mp}\neq 0$. We assume that this is always the case, which implies that only the total energy and momentum, given by summing over chiralities, are conserved. In contrast, we assume that the chiral and electric charge are separately conserved. With these assumptions in place, requiring $I_{\rm ee}$ to vanish makes the distribution functions take an LTE form with a common temperature and velocity, but distinct chemical potentials $\mu_\lambda=\mu+\lambda\mu_5$,
\begin{equation}
    f_{\vect{p},\lambda}=\frac{1}{1+e^{(\epsilon_p-\vect{u}\cdot\vect{p}-\mu_\lambda)/T}}~.
\end{equation}
From here, we can compute the total energy, charge and axial charge in LTE to be 
\begin{subequations}\label{eqn:two_currents_thermodynamics}
\begin{align}
    \epsilon&=\sum_\lambda\sum_{p,h}\int\frac{\dif^3\vect{p}}{(2\pi)^3}\epsilon_p f_{\vect{p},\lambda}=\frac{7\pi^2T^4}{60}+\frac{T^2\left(\mu^2+\mu_5^2\right)}{2}+\frac{\mu^4+6\mu^2\mu_5^2+\mu_5^4}{4\pi^2}~,\\
    n&=\sum_\lambda\sum_{p,h}\int\frac{\dif^3\vect{p}}{(2\pi)^3}f_{\vect{p},\lambda}=\frac{\mu\left(\pi^2T^2+\mu^2+3\mu_5^2\right)}{3\pi^2}~,\\
    n_5&=\sum_\lambda\lambda\sum_{p,h}\int\frac{\dif^3\vect{p}}{(2\pi)^3}f_{\vect{p},\lambda}=\frac{\mu_5\left(\pi^2T^2+3\mu^2+\mu_5^2\right)}{3\pi^2}~.
\end{align}    
\end{subequations}
The first sum is over the two chiralities (the axial quantities are weighted by $\lambda$), the second sum is over particle and hole contributions. We have assumed that the two species have the same linear dispersion relation $\epsilon_p=p$ as happens for Type I Weyl semimetals. These quantities \eqref{eqn:two_currents_thermodynamics} obey the thermodynamic relation
\begin{align}
    P+\epsilon=sT+n\mu+n_5\mu_5
\end{align}
where $P=\epsilon/3$ and $s=\frac{\partial P}{\partial T}$.

We can once again consider introducing relaxation into \eqref{eq:ChiralBoltz}. This can be performed as in section \ref{sec:SingleCurrent}, by the addition of our novel RTA to the total collision integral. In particular, we can very easily introduce momentum relaxation when we assume that elastic scattering with impurities cannot change the chirality. In that case, including the collision integral  \eqref{eqn:momentum_relaxation} in the non-mixing terms of the total collision integral (one for each chirality) we find the momentum relaxation terms we saw in \ref{sec:relaxing_normal_charge} and as they were first found in \cite{gorbar:consistenthydrodynamics}.

As far as relaxing the rest of the charges is concerned, there are several possible modifications of the RTA we can use: 

\paragraph{Intra-valley scattering}

We start with the simplest RTA, that does not mix chiralities, namely
\begin{equation}\label{eqn:intra_valley_RTA}
    I_\text{coll}=-\frac{f_\lambda^{(0)}-\bar f_\lambda^{(0)}}{\tau}~,
\end{equation}
where again $f^{(0)}_\lambda$ is the LTE at zero velocity and the bar means global thermodynamic equilibrium. This collision integral relaxes all three charges at the same rate $\tau$, which is indeed one of the possible solutions to the Onsager constraints \eqref{eqn:Onsager_constraints} - see section~\ref{sec:relaxing_normal_charge}. From a physical perspective, this kind of RTA mimics the intra-valley scattering, i.e.~scattering within each Weyl cone. Microscopically, this collision integral should have a justification similar to the single current case: electrons of a specific Weyl node interact with phonons without scattering to the other node.

As before, we can allow for a generic energy dependence of the relaxation time $\tau=\tau(\epsilon_p)$ and again find generalized relaxations upon linearization. The resulting equations satisfy Onsager reciprocity identically.

\paragraph{Inter-valley scattering}
In the presence of two particles species, we can also consider more generic collision integrals. In particular, we can consider impurity-mediated inter-valley scattering. These RTA are usually implemented out of equilibrium (away from the hydrodynamic regime) by subtracting from $f_{\vect{p}}$ its average over the angular directions, see e.g. \cite{dantas:magnetotransportmultiweyl,zyuzin:magnetotransport}. In the present case the $f^{(0)}_\lambda$ are isotropic and we must follow a different route, but we can nonetheless write an expression resembling inter-valley scattering
\begin{equation}\label{eqn:inter_valley_RTA}
    I_\text{coll}=-\frac{f^{(0)}_\lambda-f^{(0)}_{-\lambda}}{2\tau}~.
\end{equation}
This inter-valley RTA can be argued \cite{lucas:hydrodynamictheory} to follow from an impurity-mediated inter-valley scattering from a collision integral of the form
\begin{equation}
    I_\text{$\lambda$,i}=\int\dif^3\vect{p}'W_{\vect{p}\rightarrow\vect{p}'}[f_{\vect{p},\lambda}-f_{\vect{p}',-\lambda}]\delta(\epsilon_p-\epsilon_{p'})~.
\end{equation}

The above RTA \eqref{eqn:inter_valley_RTA}, when $\tau$ is constant, conserves energy and charge, while it relaxes axial charge to zero. This means only the chiral charge equation is modified to
\begin{equation}
    \partial_t n_5+\dots=-\frac{n_5}{\tau}~.
\end{equation}
We thus see that the collision integral \eqref{eqn:inter_valley_RTA} tries to destroy any imbalance between the left and right components. This is essentially put into the RTA by hand, since the collision integral for inter-valley scattering vanishes only when the densities for both chiralities are equal to each other. This forces the chiral chemical potential and, hence, the chiral charge density to zero. 

One might naively believe that \eqref{eqn:inter_valley_RTA} fails the Onsager relations test as the Onsager relations \eqref{eqn:Onsager_constraints} tell us that when $\tau_{n_5n_5}^{-1}\neq 0$, we should also relax energy and charge at the same rate around a generic equilibrium. In our case, however, we are not around a generic equilibrium, but around an equilibrium with $n_5=\mu_5=0$. This means that we must apply our Onsager constraints in the same limit (remember that we are linearising in hydrodynamics around an equilibrium state that sets to zero all the collision integrals). If we do this, we indeed find that time-reversal invariance is preserved in this case as well.

Going further and allowing for a generic energy dependence of $\tau$ in \eqref{eqn:inter_valley_RTA} once again induces generalised relaxations in the equations of motion but only for the chiral charge conservation equation. Once again, the linearized equations obey Onsager reciprocity in the limit $\mu_5=0$. 

{\ }

As we noted above, electric charge is not conserved by the intra-valley scattering RTA \eqref{eqn:intra_valley_RTA} similarly to what we found for the single current model. We can again resolve this issue in exactly the same way as we did at the end of section \ref{sec:SingleCurrent}; we can impose constraints on the microscopic times $\tau_{j}$ such that charge is conserved ($\tau_{nn}^{-1}=\tau_{n\epsilon}^{-1}=\tau_{nn_5}^{-1}=0$). We thus have three equations which need to be solved in terms of an arbitrary number of $\tau_{j}$. As before we cannot always guarantee a (non-fine tuned) solution exists, but otherwise we do not expect any issue with enforcing charge conservation. In order to avoid the issue of fine tuning altogether, we can search for an alternate modified RTA, that conserves charge identically. We discuss this alternate theory, also referred to as the BKG model  \cite{PhysRev.94.511,denicol:novelRTA,cercignani:mathematicalmethods} in the next section.

\subsection{A priori conserved charge: The BKG model}
\label{sec:BKG}

In this section, we apply our RTA approach to a particular kinetic theory model, the BKG model \cite{PhysRev.94.511,denicol:novelRTA,cercignani:mathematicalmethods}. The advantage of the BKG model over other RTAs is the fact that it enforces charge conservation at the level of the Boltzmann equation, i.e. it leads to relaxation times that automatically respect Onsager reciprocity even when the charge relaxation time vanishes.

The BKG model is usually employed to simplify the electron-electron collision integral, while maintaining some important properties. First we express the distribution function in terms of an auxiliary function $h_{\vect{p}}$,
\begin{equation}
\label{eq:DefOfh}
	f_{\vect{p}}=f^{(0)}+\delta f_{\vect{p}}=f^{(0)}(1+h_{\vect{p}})~.
\end{equation}
Subsequently, we linearize the electron-electron collision integral in $h_{\vect{p}}$ such that
\begin{equation}
\label{Eq:Linearization}
I_\text{ee} \simeq L_\text{ee}h_{\vect{p}} + \mathcal{O}^{2}(\delta) ~,
\end{equation}
where $L_\text{ee}$ is a linear operator acting on $h_{\vect{p}}$. Explicit derivations show that $L_\text{ee}$ has three zero modes
\begin{equation}
\label{Eq:ZeroModes}
L_\text{ee}1=0~~,~~ L_\text{ee}\vect{p}=0~~,~~ L_\text{ee}\epsilon_{\vect{p}}=0~,
\end{equation}
which upon integrating over momentum space lead to the conservation of charge, momentum and energy, respectively. These properties \eqref{Eq:ZeroModes} are not shared by the widely employed standard RTA approximation of $L_\text{ee}$ where one approximates
\begin{equation}\label{Eq:linearizedRTA_ee}
	L_\text{ee}h_{\vect{p}}\approx L_\text{RTA}h_{\vect{p}}=-f^{(0)}\frac{h_{\vect{p}}}{\tau}~.
\end{equation}
The BKG model consists of modifying this naive $L_{\rm RTA}$ so that charge, energy and momentum conservation are restored. This fact was used in \cite{denicol:novelRTA} to derive first order hydrodynamics in a generic hydrodynamic frame.

The procedure for motivating the BKG model goes thusly: within this linearization approximation of \eqref{Eq:Linearization}, we may think of $L_\text{ee}$ as a linear operator in the Hilbert space spanned by the real functions $h_{\vect{p}}$, whose inner product is defined via
\begin{equation}
	(h,g)=\int\dif^3\vect{p}\ f^{(0)}h_{\vect{p}}g_{\vect{p}}~.
\end{equation}
With respect to this inner product, $L_\text{ee}$ is Hermitian and negative-semidefinite
\begin{equation}
	(g,L_\text{ee}h)=(L_\text{ee}g,h) ~~,~~	(h,L_\text{ee}h)\leq 0~,
\end{equation}
where the inequality is saturated when $h$ is a zero mode (conserved quantity), $L_\text{ee}h=0$. From this, we see that the $L_\text{RTA}$ in \eqref{Eq:linearizedRTA_ee} is proportional to the identity in this Hilbert space. We can therefore seek another RTA-like operator which differs from the standard one \eqref{Eq:linearizedRTA_ee} such that it is automatically zero on the quantities we want to conserve. We allow ourselves to decompose the identity operator into the orthonormal eigenmodes of $L_\text{ee}$, denoted $\psi^a$. Such a resolution contains the projector to the zero modes of $L_\text{ee}$, which we can express as 
\begin{equation}
\label{Eq:Projector}
P^{(0)} = \sum_{L_\text{ee}\psi^b = 0}  \psi^b\tilde{\psi}^b~, 
\end{equation}  
where $\tilde{\psi}^b$ is thought of as a linear functional on the $\psi$s (a bra in Dirac's bra-ket notation) defined via $ (\psi^a,  \psi^b) =\tilde{\psi}^a(\psi^b)$.\footnote{ We assume the $\psi^b$ form an orthonormal basis for the degenerate subspace of $L$.} The BKG model then amounts to removing the projector to the zero modes one wants to keep conserved.

We can now follow the same reasoning for other collision integrals. In particular, consider the case of linearized electron-phonon interactions $L_\text{ep}$. This time the linearization is around the global equilibrium state defined by $\bar{f}^{(0)}$ 
 \begin{equation}
 	f^{(0)}=\bar{f}^{(0)}+\delta f=\bar{f}^{(0)}(1+h)~.
 \end{equation}
 and the inner product is given by
 \begin{equation}
 	(h,g)=\langle hg\rangle_{\bar{0}}=\int\dif^3\vect{p}\ \bar{f}^{(0)}hg~.
 \end{equation}
 The $I_\text{ep}$ collision integral does not conserve energy or momentum, hence we expect only one zero mode associated with charge conservation. As we saw in section \ref{sec:KinT}, such a zero mode is missing from the equivalent, standard RTA 
\begin{equation}
	\label{Eq:linearizedRTA}
	L_\text{ep}h\approx L_\text{RTA}h=-\bar{f}^{(0)}\frac{h}{\tau}~.
\end{equation}
As such, we must first consider a generic RTA operator of the form 
\begin{equation}
\label{Eq:GenericBKG}
L_\text{ep} \approx L_{*}= -{\bar{f}^{(0)}\over \tau}\sum_{i,j}a_{i,j}\psi^i\tilde{\psi}^j =  -{\bar{f}^{(0)}\over \tau}\left[1+ \sum_{i,j}(a_{i,j}-\delta_{ij})\psi^i\tilde{\psi}^j\right]~.
\end{equation}
where $a_{i,j}$ is a matrix of momentum-independent coefficients. This amounts to assuming that the full linear Boltzmann operator $L_\text{ep}$ and the RTA operator are no longer diagonal in the same basis. This is equivalent to the the process described in section \ref{sec:SingleCurrent} to move the equilibrium from a generic LTE $f^{(0)}$ to the global equilibrium $\bar{f}^{(0)}$. 
  
We now impose that $L_{*}$ of \eqref{Eq:GenericBKG} should conserve charge identically, i.e. we must ensure that $(\psi^1, L_{*} h) = 0$ for any function $h$, where $\psi^1$ denotes the charge zero mode of $L_\text{ep}$. We achieve this by setting $a_{1,i}$ of \eqref{Eq:GenericBKG} to zero. However $a_{i,1}$ (note the order of indices) has no such constraint. This allows for $\psi^1$ to not be a zero mode of $L_{*}$, resulting in a non-conserved phase space charge density. We consider this an indication that a system described by $L_{*}$ is weakly open. This is further confirmed by the fact that $L_{*}$ may not enjoy any other property of the full linearized collision operator, such as Hermiticity or being negative semi-definite. Consequently, except for electric charge, it might not conserve any other charge or obey the second law of thermodynamics and the associated relaxation times stemming from $L_{*}$ need not be positive.

For an initial foray, using all the modes of $L_\text{ep}$ in $L_{*}$ seems excessive since we are only interested in the first moments of the conserved quantities. So, we restrict $a_{i, j>5}=0=a_{i>5,j}$, where we assume that $i,j = 1,2,3,4,5$ denote the modes related to charge, energy and momentum conservation respectively. That is $\psi^1 \sim 1$, $\psi^2 \sim \epsilon_p$ and $\psi^i \sim p^{i-2},~i=3,4,5$ up to orthonormalization. Furthermore, we can decouple charge and energy relaxation from momentum relaxation by requiring $a_{i,j}$ to take a block diagonal form. Doing so allows treating momentum relaxation independently from energy and charge relaxation as in section \ref{sec:generalised relaxations}. Thus, we assume that $a_{i,j}$ is simply a $2\times 2$ matrix which reads
\begin{equation}
a_{i,j} = \begin{pmatrix}
0 & 0
\\
\alpha_2 & a_1
\end{pmatrix}
\Rightarrow \alpha_{i,j} \equiv a_{i,j}-\delta_{ij} = 
\begin{pmatrix}
-1 & 0
\\
\alpha_2 & \alpha_1
\end{pmatrix}~.
\end{equation}
To show this explicitly, we construct an orthonormal basis out of $1$ and $\epsilon_{\vect{p}}$, which reads
\begin{align}
\label{eq:ModesOrtho}
\psi^{1} = {1\over \sqrt{\bar{n}}}\quad,\qquad\psi^2 = {\bar{\epsilon}-\bar{n}\epsilon_p\over \sqrt{\bar{n}^2\bar{\epsilon^2}-\bar{\epsilon}^2\bar{n}}}~. 
\end{align}  
This leads to an $L_{*}$ of the form
\begin{equation}
L_{*} = -{\bar{f}^{(0)}\over \tau}\sum_{i,j}\left(\alpha_2 \psi^2\tilde{\psi}^1 + a_1 \psi^2\tilde{\psi^2}\right) =  -{\bar{f}^{(0)}\over \tau}\left[1-\psi^1\tilde{\psi}^1 + \alpha_2 \psi^2\tilde{\psi}^1 + \alpha_1 \psi^2\tilde{\psi^2}\right]~.
\end{equation}
When $\alpha_2 = 0$ and $\alpha_1 = -1$ we recover the BKG model used for electron-electron collisions. These two extra parameters, while ad hoc, give us enough freedom to make the relaxation times of the linearized hydrodynamic equations satisfy Onsager reciprocity. Upon substituting $h$ in terms of Boltzmann distributions via \eqref{eq:DefOfh} and the $\psi$s via \eqref{eq:ModesOrtho}, we find that the linearized Boltzmann equation becomes
\begin{equation}
\label{eq:BKGModified}
\partial_t f^{(0)} + \dots = -{1 \over \tau}\left[f^{(0)} - {n \over \bar{n}}\bar{f}^{(0)} + \tilde{\alpha}_2 \bar{f}^{(0)} (n-\bar{n}) + \tilde{\alpha}_1 \bar{f}^{(0)}(\bar{\epsilon}n-\bar{n}\epsilon)\right]~,
\end{equation}
where
\begin{equation}
\tilde{\alpha}_1 = \alpha_1 {\bar{\epsilon}-\bar{n}\epsilon_p\over \bar{n}^2\bar{\epsilon^2} - \bar{\epsilon}^2\bar{n}}
~~,~~
\tilde{\alpha}_2 = \alpha_2 {\bar{\epsilon}-\bar{n}\epsilon_p\over \sqrt{\bar{n}^3\bar{\epsilon^2} - \bar{\epsilon}^2\bar{n}^2}} ~.
\end{equation}
Note that $\tilde{\alpha}_1$ and $\tilde{\alpha}_2$ are functions of momenta, in contrast to $\alpha_1$ and $\alpha_2$.\footnote{Of course, $\alpha_{1}$ and $\alpha_2$ maybe functions of temperature and chemical potential.} In this way, \eqref{eq:BKGModified} resembles the expansion in the energy-dependent relaxation time \eqref{eqn:generalised_relaxations_collision_integral}. Enforcing the right-hand side of \eqref{eq:BKGModified} to vanish identically whenever $f^{(0)} = \bar{f}^{(0)}$ leads to the constraints $n = \bar{n}$ and $\epsilon = \bar{\epsilon}$, i.e. it fixes the equilibrium charge densities to take their expected form. 

We can now derive the hydrodynamic equations of motion. Charge is identically conserved, as can be seen by integrating \eqref{eq:BKGModified} over momentum,\footnote{It is easy to show $\langle\tilde{\alpha}_2\rangle_{\bar{0}} = 0 = \langle\tilde{\alpha}_1\rangle_{\bar{0}}$.} while the energy equation reads
\begin{equation}
	\partial_t\epsilon+\dots=-\frac{\epsilon-\bar\epsilon}{\tau}+\frac{\bar\epsilon}{\bar n}\frac{n-\bar n}{\tau}-\frac{\alpha_2(\bar\epsilon^2-\bar n\bar{\epsilon^2})}{\sqrt{\bar{n}^3\bar{\epsilon^2} - \bar{\epsilon}^2\bar{n}^2}}\frac{n-\bar n}{\tau} \\
	-\frac{\alpha_1(\bar\epsilon^2-\bar n\bar{\epsilon^2})}{\bar n^2\bar{\epsilon^2}-\bar n\bar\epsilon^2}\frac{\bar\epsilon n-\bar n\epsilon}{\tau}
\end{equation}
Upon linearization, we obtain two relaxation times that depend on either $\alpha_1$ or $\alpha_2$. Thus we can fix the relaxation times to take Onsager reciprocal values at will. Note that this is also possible if at least one of the $\alpha_i$ is non-zero. It would be interesting to understand whether the values of $\alpha_1$ and $\alpha_2$ can be constrained from fundamental principles or they should be regarded as phenomenological parameters.

We thus see that the novel RTA we introduced in \ref{sec:SingleCurrent} and which leads to generalized relaxation times can also account for an identically conserved charge. This is achieved via two possible independent mechanisms, i) the introduction of phase space interactions between the energy and charge modes of the linearized Boltzmann operator, $\alpha_2 \neq 0$ and/or ii) the violation of energy conservation $\alpha_1 \neq -1$.

\section{Conclusions}
\label{sec:Conclusions}

There are two main conclusions arising from this paper. First, generalized relaxations are necessary for a hydrodynamics of anomalous fluids consistent with Onsager reciprocity, charge conservation and phenomenological constraints on the fluid's DC conductivities. This, however, always comes at the expense of violating the second law of thermodynamics. A system exhibiting generalized relaxations is necessarily open. The relaxation times themselves are not arbitrary, but fall into certain families of solutions, some of which where presented explicitly in section \ref{sec:generalised relaxations}.

The second main conclusion of our paper is that these generalized relaxations can be derived directly from kinetic theory, upon using a novel RTA. In particular, our RTA works at zero order in the hydrodynamic derivative expansion and presumes that the relaxation times themselves are energy-dependent and pick out their ``preferred'' equilibrium distribution function $\bar{f}^{(0)}$. We have shown that the families of relaxation times found in \ref{sec:generalised relaxations} can be found directly from microscopic considerations. In addition, charge conservation may or may not be imposed a priori, depending on whether the RTA of section \ref{sec:SingleCurrent} or of section \ref{sec:BKG} is used, respectively.

There are several possible directions that open up based on our results; from the purely hydrodynamic point of view, we can ask ourselves what happens to the Ward identities obeyed by the hydrodynamic Green's functions in the presence of generalized relaxations. Does the Wiedermann-Franz law change because of the dependence of the DC conductivities on the generalized relaxation times? Additionally, we should also understand whether the discrepancy on the magnetoconductivities of WSMs we found in \ref{sec:generalised relaxations} can be remedied or certified by modifying our theory of hydrodynamics or from direct experimental observation. 

From the kinetic theory point of view, we must understand whether our prescription for obtaining generalized relaxation times works for spectra besides the linear one of WSMs, for suitable non-analytic energy-dependence of the microscopic relaxation time $\tau(\epsilon_p)$, or choices of collision integral (i.e. via a mechanism not fixing the reference distribution $\bar{f}^{(0)}$ by hand). In addition, we should confirm from direct computation that there exist relaxation times that exhibit the energy-dependence we have assumed in section \ref{sec:KinT}. Furthermore, we should investigate whether it is possible to modify the inter-valley RTA \eqref{eqn:inter_valley_RTA} such that the WSM fluid relaxes to an equilibrium with $\mu_5 \neq 0$ and a non-trivial chiral charge density $n_5$. It might also be interesting to understand the effect of generalised relaxations for other particle statistics and dimensions, such as two spatial dimensions where interesting additional phenomena occur such as non-integer spin (see \cite{Brattan:2013wya,Brattan:2014moa} and references therein).
 Finally, as we mentioned, it would be interesting to understand whether the arbitrary coefficients in our charge-preserving RTA of section \ref{sec:BKG} can be constrained by additional fundamental principles besides Onsager reciprocity, such as linear stability, which could provide some bounds for the parameters.
 
 \acknowledgments
 
A.A. and I.M. have been partially supported by the “Curiosity Driven Grant 2020” of the University of Genoa and the INFN Scientific Initiative SFT: “Statistical Field Theory, Low-Dimensional Systems, Integrable Models and Applications”. This project has also received funding from the European Union’s Horizon 2020 research and innovation programme under the Marie Sklodowska-Curie grant agreement No. 101030915.

\bibliography{ref}

\providecommand{\href}[2]{#2}\begingroup\raggedright\begin{thebibliography}{10}

\bibitem{armitage:weylanddirac}
N.~P. Armitage, E.~J. Mele, and A.~Vishwanath, {\it Weyl and dirac semimetals
  in three-dimensional solids},  {\em Rev. Mod. Phys.} {\bf 90} (Jan, 2018)
  015001.

\bibitem{burkov:weylmetals}
A.~Burkov, {\it Weyl metals},  {\em Annual Review of Condensed Matter Physics}
  {\bf 9} (mar, 2018) 359--378.

\bibitem{Hosur_2013}
P.~Hosur and X.~Qi, {\it Recent developments in transport phenomena in weyl
  semimetals},  {\em Comptes Rendus Physique} {\bf 14} (nov, 2013) 857--870.

\bibitem{Nielsen:1980rz}
H.~B. Nielsen and M.~Ninomiya, {\it {Absence of Neutrinos on a Lattice. 1.
  Proof by Homotopy Theory}},  {\em Nucl. Phys. B} {\bf 185} (1981) 20.
  [Erratum: Nucl.Phys.B 195, 541 (1982)].

\bibitem{NIELSEN1981219}
H.~Nielsen and M.~Ninomiya, {\it A no-go theorem for regularizing chiral
  fermions},  {\em Physics Letters B} {\bf 105} (1981), no.~2 219--223.

\bibitem{PhysRev.177.2426}
S.~L. Adler, {\it Axial-vector vertex in spinor electrodynamics},  {\em Phys.
  Rev.} {\bf 177} (Jan, 1969) 2426--2438.

\bibitem{Bell1969}
J.~S. Bell and R.~Jackiw, {\it A pcac puzzle:
  $\pi$0{\textrightarrow}$\gamma$$\gamma$ in the $\sigma$-model},  {\em Il
  Nuovo Cimento A (1965-1970)} {\bf 60} (Mar, 1969) 47--61.

\bibitem{nielsen:adler-bell-jackiw}
H.~B. Nielsen and M.~Ninomiya, {\it {ADLER-BELL-JACKIW ANOMALY AND WEYL
  FERMIONS IN CRYSTAL}},  {\em Phys. Lett. B} {\bf 130} (1983) 389--396.

\bibitem{landsteiner:notesanomaly}
K.~Landsteiner, {\it Notes on {{Anomaly Induced Transport}}},  {\em Acta Phys.
  Pol. B} {\bf 47} (2016), no.~12 2617,
  [\href{http://arxiv.org/abs/1610.04413}{{\tt arXiv:1610.04413}}].

\bibitem{landsteiner:negativemagnetoresistivity}
K.~Landsteiner, Y.~Liu, and Y.-W. Sun, {\it Negative magnetoresistivity in
  chiral fluids and holography},  {\em J. High Energ. Phys.} {\bf 2015} (Mar.,
  2015) 127.

\bibitem{landsteiner:anomaloustransport}
K.~Landsteiner, {\it Anomalous transport of weyl fermions in weyl semimetals},
  {\em Phys. Rev. B} {\bf 89} (Feb, 2014) 075124.

\bibitem{chernodub:thermaltransport}
M.~N. Chernodub, Y.~Ferreiros, A.~G. Grushin, K.~Landsteiner, and M.~A.~H.
  Vozmediano, {\it {Thermal transport, geometry, and anomalies}},  {\em Phys.
  Rept.} {\bf 977} (2022) 1--58, [\href{http://arxiv.org/abs/2110.05471}{{\tt
  arXiv:2110.05471}}].

\bibitem{doi:10.1080/23746149.2017.1327329}
S.~Wang, B.-C. Lin, A.-Q. Wang, D.-P. Yu, and Z.-M. Liao, {\it Quantum
  transport in dirac and weyl semimetals: a review},  {\em Advances in Physics:
  X} {\bf 2} (2017), no.~3 518--544,
  [\href{http://arxiv.org/abs/https://doi.org/10.1080/23746149.2017.1327329}{{\tt
  https://doi.org/10.1080/23746149.2017.1327329}}].

\bibitem{sukhachov:anomalousgurzhi}
P.~O. Sukhachov and B.~Trauzettel, {\it Anomalous gurzhi effect},  {\em Phys.
  Rev. B} {\bf 105} (Feb, 2022) 085141.

\bibitem{lucas:hydrodynamictheory}
A.~Lucas, R.~A. Davison, and S.~Sachdev, {\it Hydrodynamic theory of
  thermoelectric transport and negative magnetoresistance in {{Weyl}}
  semimetals},  {\em Proc. Natl. Acad. Sci. U.S.A.} {\bf 113} (Aug., 2016)
  9463--9468, [\href{http://arxiv.org/abs/1604.08598}{{\tt arXiv:1604.08598}}].

\bibitem{abbasi:magnetotransportanomalous}
N.~Abbasi, A.~Ghazi, F.~Taghinavaz, and O.~Tavakol, {\it Magneto-transport in
  an anomalous fluid with weakly broken symmetries, in weak and strong regime},
   {\em J. High Energ. Phys.} {\bf 2019} (May, 2019) 206.

\bibitem{abbasi:magnetotransportchiral}
N.~Abbasi, F.~Taghinavaz, and O.~Tavakol, {\it Magneto-transport in a chiral
  fluid from kinetic theory},  {\em J. High Energ. Phys.} {\bf 2019} (Mar.,
  2019) 51.

\bibitem{gorbar:nonlocaltransport}
E.~V. Gorbar, V.~A. Miransky, I.~A. Shovkovy, and P.~O. Sukhachov, {\it
  Nonlocal transport in weyl semimetals in the hydrodynamic regime},  {\em
  Phys. Rev. B} {\bf 98} (Jul, 2018) 035121.

\bibitem{son:chiralanomaly}
D.~T. Son and B.~Z. Spivak, {\it Chiral anomaly and classical negative
  magnetoresistance of weyl metals},  {\em Phys. Rev. B} {\bf 88} (Sep, 2013)
  104412.

\bibitem{Nag_2021}
T.~Nag and S.~Nandy, {\it Magneto-transport phenomena of type-i multi-weyl
  semimetals in co-planar setups},  {\em Journal of Physics: Condensed Matter}
  {\bf 33} (nov, 2020) 075504.

\bibitem{Gooth_2018}
J.~Gooth, F.~Menges, N.~Kumar, V.~Sü{\ss}, C.~Shekhar, Y.~Sun, U.~Drechsler,
  R.~Zierold, C.~Felser, and B.~Gotsmann, {\it Thermal and electrical
  signatures of a hydrodynamic electron fluid in tungsten diphosphide},  {\em
  Nature Communications} {\bf 9} (oct, 2018).

\bibitem{Vu:thermal_chiral_anomaly}
D.~Vu, W.~Zhang, C.~{\c{S}}ahin, M.~E. Flatt{\'{e}}, N.~Trivedi, and J.~P.
  Heremans, {\it Thermal chiral anomaly in the magnetic-field-induced ideal
  weyl phase of bi1-{xSbx}},  {\em Nature Materials} {\bf 20} (jun, 2021)
  1525--1531.

\bibitem{Xiong:evidence_chiral_anomaly}
J.~Xiong, S.~K. Kushwaha, T.~Liang, J.~W. Krizan, M.~Hirschberger, W.~Wang,
  R.~J. Cava, and N.~P. Ong, {\it Evidence for the chiral anomaly in the dirac
  semimetal na<sub>3</sub>bi},  {\em Science} {\bf 350} (2015), no.~6259
  413--416,
  [\href{http://arxiv.org/abs/https://www.science.org/doi/pdf/10.1126/science.aac6089}{{\tt
  https://www.science.org/doi/pdf/10.1126/science.aac6089}}].

\bibitem{PhysRevX.5.031023}
X.~Huang, L.~Zhao, Y.~Long, P.~Wang, D.~Chen, Z.~Yang, H.~Liang, M.~Xue,
  H.~Weng, Z.~Fang, X.~Dai, and G.~Chen, {\it Observation of the
  chiral-anomaly-induced negative magnetoresistance in 3d weyl semimetal taas},
   {\em Phys. Rev. X} {\bf 5} (Aug, 2015) 031023.

\bibitem{Gooth_2017}
J.~Gooth, A.~C. Niemann, T.~Meng, A.~G. Grushin, K.~Landsteiner, B.~Gotsmann,
  F.~Menges, M.~Schmidt, C.~Shekhar, V.~Sü{\ss}, R.~Hühne, B.~Rellinghaus,
  C.~Felser, B.~Yan, and K.~Nielsch, {\it Experimental signatures of the mixed
  axial{\textendash}gravitational anomaly in the weyl semimetal {NbP}},  {\em
  Nature} {\bf 547} (jul, 2017) 324--327.

\bibitem{Jia2016ThermoelectricSO}
Z.~Jia, C.-Z. Li, X.~Li, J.~ren Shi, Z.~Liao, D.~Yu, and X.~Wu, {\it
  Thermoelectric signature of the chiral anomaly in cd3as2},  {\em Nature
  Communications} {\bf 7} (2016).

\bibitem{amoretti:framedependence}
A.~Amoretti, D.~K. Brattan, L.~Martinoia, and I.~Matthaiakakis, {\it {Leading
  order magnetic field dependence of conductivities in anomalous
  hydrodynamics}},  {\em Phys. Rev. D} {\bf 108} (2023), no.~1 016003,
  [\href{http://arxiv.org/abs/2212.09761}{{\tt arXiv:2212.09761}}].

\bibitem{amoretti:restoringtime}
A.~Amoretti, D.~K. Brattan, L.~Martinoia, and I.~Matthaiakakis, {\it {Restoring
  time-reversal covariance in relaxed hydrodynamics}},  {\em Phys. Rev. D} {\bf
  108} (2023), no.~5 056003, [\href{http://arxiv.org/abs/2304.01248}{{\tt
  arXiv:2304.01248}}].

\bibitem{gorbar:consistenthydrodynamics}
E.~V. Gorbar, V.~A. Miransky, I.~A. Shovkovy, and P.~O. Sukhachov, {\it
  Consistent hydrodynamic theory of chiral electrons in weyl semimetals},  {\em
  Phys. Rev. B} {\bf 97} (Mar, 2018) 121105.

\bibitem{dantas:magnetotransportmultiweyl}
R.~M.~A. Dantas, F.~{Pe{\~n}a-Benitez}, B.~Roy, and P.~Sur{\'o}wka, {\it
  Magnetotransport in multi-{{Weyl}} semimetals: A kinetic theory approach},
  {\em J. High Energ. Phys.} {\bf 2018} (Dec., 2018) 69.

\bibitem{kovtun:lecturenotes}
P.~Kovtun, {\it {Lectures on hydrodynamic fluctuations in relativistic
  theories}},  {\em J. Phys. A} {\bf 45} (2012) 473001,
  [\href{http://arxiv.org/abs/1205.5040}{{\tt arXiv:1205.5040}}].

\bibitem{Amoretti:2022acb}
A.~Amoretti and D.~K. Brattan, {\it {On the hydrodynamics of (2 +
  1)-dimensional strongly coupled relativistic theories in an external magnetic
  field}},  {\em Mod. Phys. Lett. A} {\bf 37} (2022), no.~21 2230010,
  [\href{http://arxiv.org/abs/2209.11589}{{\tt arXiv:2209.11589}}].

\bibitem{Amoretti:2022ovc}
A.~Amoretti, D.~K. Brattan, L.~Martinoia, and I.~Matthaiakakis, {\it
  {Non-dissipative electrically driven fluids}},  {\em JHEP} {\bf 05} (2023)
  218, [\href{http://arxiv.org/abs/2211.05791}{{\tt arXiv:2211.05791}}].

\bibitem{rogatko:magnetotransport}
M.~Rogatko and K.~I. Wysokinski, {\it {Magnetotransport of Weyl semimetals with
  $\mathbb{Z}_2$ topological charge and chiral anomaly}},  {\em JHEP} {\bf 01}
  (2019) 049, [\href{http://arxiv.org/abs/1810.07521}{{\tt arXiv:1810.07521}}].

\bibitem{sadofyev:dragsuppression}
A.~V. Sadofyev and Y.~Yin, {\it Drag suppression in anomalous chiral media},
  {\em Phys. Rev. D} {\bf 93} (June, 2016) 125026.

\bibitem{stephanov:nodragframe}
M.~A. Stephanov and H.-U. Yee, {\it No-{{Drag Frame}} for {{Anomalous Chiral
  Fluid}}},  {\em Phys. Rev. Lett.} {\bf 116} (Mar., 2016) 122302.

\bibitem{PhysRevResearch.2.013088}
K.~Das and A.~Agarwal, {\it Thermal and gravitational chiral anomaly induced
  magneto-transport in weyl semimetals},  {\em Phys. Rev. Res.} {\bf 2} (Jan,
  2020) 013088.

\bibitem{PhysRevB.93.085107}
B.~Z. Spivak and A.~V. Andreev, {\it Magnetotransport phenomena related to the
  chiral anomaly in weyl semimetals},  {\em Phys. Rev. B} {\bf 93} (Feb, 2016)
  085107.

\bibitem{sharma:nernst_magnetothermal}
G.~Sharma, P.~Goswami, and S.~Tewari, {\it Nernst and magnetothermal
  conductivity in a lattice model of weyl fermions},  {\em Phys. Rev. B} {\bf
  93} (Jan, 2016) 035116.

\bibitem{lundgren:thermoelectric_properties}
R.~Lundgren, P.~Laurell, and G.~A. Fiete, {\it Thermoelectric properties of
  weyl and dirac semimetals},  {\em Phys. Rev. B} {\bf 90} (Oct, 2014) 165115.

\bibitem{narozhny:hydrodynamicapproach}
B.~N. Narozhny, {\it {Hydrodynamic approach to two-dimensional electron
  systems}},  {\em Riv. Nuovo Cim.} {\bf 45} (2022), no.~10 661--736,
  [\href{http://arxiv.org/abs/2207.10004}{{\tt arXiv:2207.10004}}].

\bibitem{huang:statisticalmechanics}
K.~Huang, {\em Statistical mechanics}.
\newblock Wiley, 1987.

\bibitem{tong:statisticalphysics}
D.~Tong, {\it Kinetic theory},  2012.

\bibitem{denicol:microscopicfoundations}
G.~S. Denicol and D.~H. Rischke, {\em Microscopic Foundations of Relativistic
  Fluid Dynamics}.
\newblock Springer Cham, 2021.

\bibitem{landau:physicalkinetics}
L.~P. Pitaevskii and E.~M. Lifshitz, {\em Physical Kinetics: Volume 10 (Course
  of Theoretical Physics)}.
\newblock Butterworth-Heinemann, 1981.

\bibitem{pal:resistivity}
H.~K. Pal, V.~I. Yudson, and D.~L. Maslov, {\it Resistivity of
  non-galilean-invariant fermi- and non-fermi liquids},  {\em Lithuanian
  Journal of Physics} {\bf 52} (2012), no.~2 142--164.

\bibitem{DASH2022137202}
D.~Dash, S.~Bhadury, S.~Jaiswal, and A.~Jaiswal, {\it Extended relaxation time
  approximation and relativistic dissipative hydrodynamics},  {\em Physics
  Letters B} {\bf 831} (2022) 137202.

\bibitem{narozhny:energyrelaxation}
B.~N. Narozhny and I.~V. Gornyi, {\it Hydrodynamic approach to electronic
  transport in graphene: Energy relaxation},  {\em Frontiers in Physics} {\bf
  9} (apr, 2021).

\bibitem{cercignani:mathematicalmethods}
C.~Cercignani, {\em Mathematical methods in kinetic theory}.
\newblock Plenum Press New York, 1990.

\bibitem{ochi:electronhole}
M.~Ochi, {\it Electron-hole dichotomy for thermoelectric transport in a
  two-valley system with strong intervalley scattering},  2023.

\bibitem{son:kinetictheory}
D.~T. Son and N.~Yamamoto, {\it Kinetic theory with berry curvature from
  quantum field theories},  {\em Phys. Rev. D} {\bf 87} (Apr, 2013) 085016.

\bibitem{gorbar:consistentchiralkinetic}
E.~V. Gorbar, V.~A. Miransky, I.~A. Shovkovy, and P.~O. Sukhachov, {\it
  {Consistent Chiral Kinetic Theory in Weyl Materials: Chiral Magnetic
  Plasmons}},  {\em Phys. Rev. Lett.} {\bf 118} (2017), no.~12 127601,
  [\href{http://arxiv.org/abs/1610.07625}{{\tt arXiv:1610.07625}}].

\bibitem{pongsangangan:hydrodynamics}
K.~Pongsangangan, T.~Ludwig, H.~T.~C. Stoof, and L.~Fritz, {\it Hydrodynamics
  of charged two-dimensional dirac systems. i. thermoelectric transport},  {\em
  Phys. Rev. B} {\bf 106} (Nov, 2022) 205126.

\bibitem{bianca:boltzmanngasmixture}
C.~Bianca and C.~Dogb{\'e}, {\it On the boltzmann gas mixture equation: Linking
  the kinetic and fluid regimes},  {\em Commun. Nonlinear Sci. Numer. Simul.}
  {\bf 29} (2015) 240--256.

\bibitem{fotakis:multicomponent}
J.~A. Fotakis, E.~Moln\'ar, H.~Niemi, C.~Greiner, and D.~H. Rischke, {\it
  Multicomponent relativistic dissipative fluid dynamics from the boltzmann
  equation},  {\em Phys. Rev. D} {\bf 106} (Aug, 2022) 036009.

\bibitem{zyuzin:magnetotransport}
V.~A. Zyuzin, {\it Magnetotransport of weyl semimetals due to the chiral
  anomaly},  {\em Phys. Rev. B} {\bf 95} (Jun, 2017) 245128.

\bibitem{PhysRev.94.511}
P.~L. Bhatnagar, E.~P. Gross, and M.~Krook, {\it A model for collision
  processes in gases. i. small amplitude processes in charged and neutral
  one-component systems},  {\em Phys. Rev.} {\bf 94} (May, 1954) 511--525.

\bibitem{denicol:novelRTA}
G.~S. Rocha, G.~S. Denicol, and J.~Noronha, {\it Novel relaxation time
  approximation to the relativistic boltzmann equation},  {\em Phys. Rev.
  Lett.} {\bf 127} (Jul, 2021) 042301.

\bibitem{Brattan:2013wya}
D.~K. Brattan and G.~Lifschytz, {\it {Holographic plasma and anyonic fluids}},
  {\em JHEP} {\bf 02} (2014) 090, [\href{http://arxiv.org/abs/1310.2610}{{\tt
  arXiv:1310.2610}}].

\bibitem{Brattan:2014moa}
D.~K. Brattan, {\it {A strongly coupled anyon material}},  {\em JHEP} {\bf 11}
  (2015) 214, [\href{http://arxiv.org/abs/1412.1489}{{\tt arXiv:1412.1489}}].

\end{thebibliography}\endgroup
\bibliographystyle{JHEP}

\end{document}